\pdfoutput=1

\documentclass[11pt]{article}

\usepackage[]{acl}

\usepackage{times}
\usepackage{latexsym}
\usepackage{enumitem}
\usepackage{subcaption}
\usepackage{xcolor}

\definecolor{emphblue}{RGB}{0,90,160}
\definecolor{emphred}{RGB}{160,40,40}

\usepackage[T1]{fontenc}

\usepackage{CJKutf8}
\newcommand{\zh}[1]{\begin{CJK}{UTF8}{gbsn}#1\end{CJK}}
\usepackage{microtype}

\usepackage{inconsolata}

\usepackage{graphicx}
\usepackage{booktabs}
\usepackage{amsmath}
\usepackage{cleveref}
\usepackage{multirow}
\usepackage{url}
\usepackage{xcolor}
\usepackage{colortbl}  
\newcommand{\corrcolor}[1]{%
  \ifdim#1pt>0.9pt\cellcolor{green!40}\textcolor{black}{#1}%
  \else\ifdim#1pt>0.7pt\cellcolor{green!40}\textcolor{black}{#1}%
  \else\ifdim#1pt>0.4pt\cellcolor{yellow!70}\textcolor{black}{#1}%
  \else\ifdim#1pt>0.1pt\cellcolor{yellow!30}\textcolor{black}{#1}%
  \else\ifdim#1pt>-0.1pt\cellcolor{orange!30}\textcolor{black}{#1}%
  \else\cellcolor{red!40}\textcolor{black}{#1}%
  \fi\fi\fi\fi\fi%
}

\graphicspath{{img/}}

\title{Beyond Translation: Cross-Cultural Meme Transcreation with \\Vision-Language Models}

 \author{Yuming Zhao \hspace{5pt} 
 Peiyi Zhang \hspace{5pt}
Oana Ignat \\
Santa Clara University - Santa Clara, USA  \\
 \textit{\{yzhao4, oignat\}@scu.edu} \\  }

\begin{document}\maketitle

\begin{abstract}
Memes are a pervasive form of online communication, yet their cultural specificity poses significant challenges for cross-cultural adaptation. We study \textit{cross-cultural meme transcreation}, a multimodal generation task that aims to preserve communicative intent and humor while adapting culture-specific references. We propose a hybrid transcreation framework based on vision–language models and introduce a large-scale bidirectional dataset of Chinese and US memes. Using both human judgments and automated evaluation, we analyze 6,315 meme pairs and assess transcreation quality across cultural directions. Our results show that current vision–language models can perform cross-cultural meme transcreation to a limited extent, but exhibit clear directional asymmetries: US$\rightarrow$Chinese transcreation consistently achieves higher quality than Chinese$\rightarrow$US. We further identify which aspects of humor and visual–textual design transfer across cultures and which remain challenging, and propose an evaluation framework for assessing cross-cultural multimodal generation. Our code and dataset are publicly available at \url{https://github.com/AIM-SCU/MemeXGen}.
\end{abstract}

\section{Introduction}
\label{sec:introduction}

Memes are a dominant form of online communication, yet they are difficult to adapt across cultural contexts. A meme that resonates with US audiences may fail among Chinese users—not due to linguistic errors, but because its humor, symbolism, or visual style does not translate culturally. While literal translation preserves surface meaning, it often fails to preserve what makes a meme effective: its intent, humor, and cultural resonance.

This challenge is often described as \textit{transcreation}~\cite{khanuja2024image}: the process of adapting content across cultures by preserving communicative intent rather than literal form. For memes, transcreation requires coordinated adaptation of both text and images, grounded in culturally specific norms, references, and aesthetic preferences. As such, \textit{meme transcreation poses a fundamental multimodal and cultural generation challenge that goes beyond standard translation or captioning}.

\begin{figure}[t]
\centering
\includegraphics[width=\columnwidth]{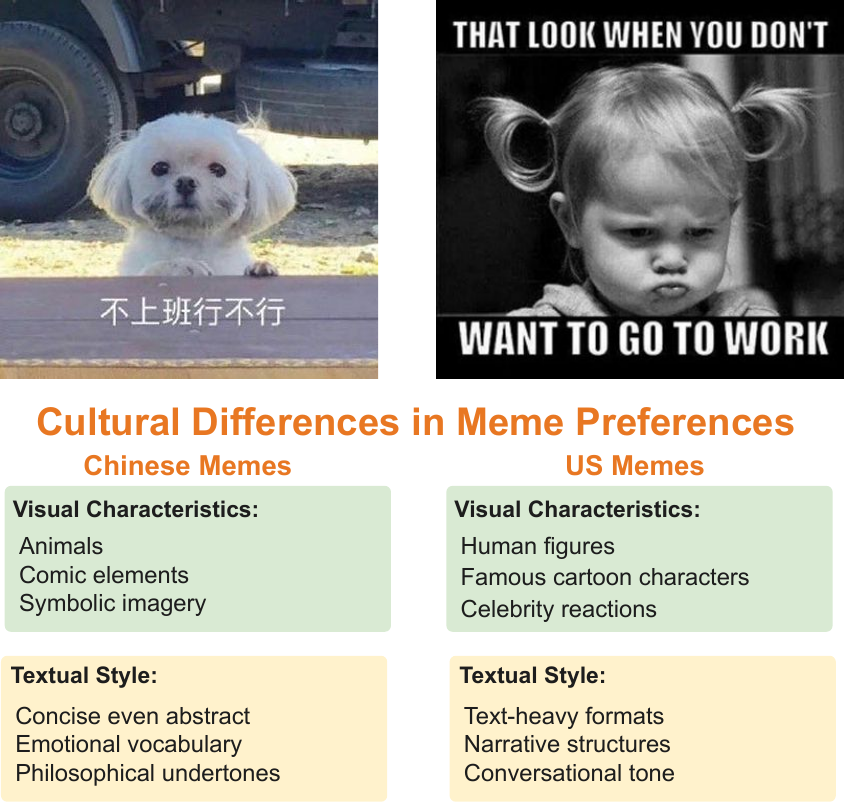}
\caption{Examples of cultural differences in meme preferences across Chinese and US contexts. Cultural preferences shape humor and visual style, creating challenges for cross-cultural meme transcreation.}
\label{fig:system_overview}
\end{figure}

\begin{table}[t]
\centering
\small
\begin{tabular}{@{}lll@{}}
\toprule
\textbf{Aspect} & \textbf{US} & \textbf{Chinese} \\
\midrule
Visual & Human, celebrity & Animal, symbolic \\
Text & Narrative, detailed & Concise, philosophical \\
Emotion & Situational & Universal \\
Humor & Sarcasm, relatable & Wordplay, cuteness \\
\bottomrule
\end{tabular}
\caption{Cross-Cultural Meme Characteristics}
\vspace{-2em}
\label{tab:cultural_differences}
\end{table}

Prior work has primarily studied memes from a recognition and analysis perspective~\cite{hazman2025whatmakes, cao2023prompthate, zhao2025memereacon}. In parallel, recent vision–language models demonstrate strong multimodal understanding capabilities. However, systematic frameworks, datasets, and evaluations for \textit{cross-cultural meme generation} remain limited. In particular, it is unclear how well current models can perform culturally grounded transcreation, how performance varies across cultural directions, and how such systems should be evaluated.

To address this gap, we empirically study cross-cultural meme transcreation between Chinese and US cultures through a hybrid framework that adapts memes while preserving communicative intent. We introduce a large-scale bidirectional dataset of original and transcreated memes and evaluate transcreation quality using both human judgments and automated evaluation. Our analysis highlights systematic directional effects and cultural factors that shape the success and limitations of current vision–language models in cross-cultural meme transcreation. Table~\ref{tab:cultural_differences} summarizes key cross-cultural distinctions considered in this work.

This paper addresses three research questions:

\begin{description}[]
    \item[\textbf{RQ1:}] How effectively can vision–language models perform cross-cultural meme transcreation while preserving intent, humor, and cultural nuance?
    \item[\textbf{RQ2:}] Does transcreation direction introduce systematic asymmetries between US$\rightarrow$Chinese and Chinese$\rightarrow$US adaptations?
    \item[\textbf{RQ3:}] How should multimodal meme transcreation be evaluated, and how do human judgments compare with automated evaluations?
\end{description}

Our Contributions.
\textbf{First, we propose a hybrid framework for meme transcreation} that balances intent preservation with cultural adaptation, offering practical guidance for culturally grounded meme adaptation.
\textbf{Second, we introduce the first bidirectional meme transcreation dataset}, containing 6,315 original memes and 6,315 corresponding transcreated memes across both Chinese$\rightarrow$US and US$\rightarrow$Chinese directions.
\textbf{Third, we present a bidirectional empirical study of Chinese and US meme transcreation}, showing that transcreation quality depends on direction and that specific aspects of internet humor—such as imagery, text style, and emotional expression—transfer unevenly across cultures.

\section{Related Work}
\label{sec:related}

\paragraph{Cultural Gaps in AI.}
Despite advances in large language models and vision–language models (VLMs), cultural gaps remain a persistent challenge~\cite{10.1609/aaai.v39i27.35092, DBLP:conf/emnlp/AdilazuardaMLSA24}. Prior work shows that NLP systems often fail to account for cross-cultural variation~\cite{hershcovich-etal-2022-challenges}, while text-to-image models tend to default to Western-centric representations~\cite{kannen2024beyond}. These biases manifest as systematic performance disparities across languages and cultures~\cite{field_survey_2021, naous2023having}. Recent benchmarks such as GlobalRG~\cite{bhatia-etal-2024-local} further highlight significant drops in VLM performance on local cultural concepts. Our work contributes to this line of research by studying explicit cultural adaptation in a generative setting, focusing on bidirectional cross-cultural transcreation.

\paragraph{Meme Understanding and Analysis.}
Most prior research on memes has focused on discriminative tasks, such as classification and detection. For example, PromptHate~\cite{cao2023prompthate} addresses hateful meme detection~\cite{pmlr-v133-kiela21a, kumar-nandakumar-2022-hate, sharma-etal-2023-characterizing}, while MGMCF~\cite{mgmcf2024} models fine-grained persuasive features~\cite{umuteam2024}. 
Other studies document systematic cultural differences in online humor~\cite{mutheu2023cross, nissenbaum2018internet, tanaka-etal-2022-learning}, analyze the sentiment of memes~\cite{sharma-etal-2020-semeval}, and show that annotators’ cultural backgrounds influence interpretation. In contrast, comparatively little work has explored \emph{generative} meme tasks, particularly those requiring culturally grounded adaptation rather than classification.

\paragraph{Cross-Cultural Transcreation.}
Transcreation has recently emerged as a framework for adapting content across cultures beyond literal translation. \citet{khanuja2024image} introduce image transcreation and show that models struggle to balance semantic preservation with cultural appropriateness, motivating dedicated evaluation metrics~\cite{khanuja2024towards}. While meme datasets such as MemeCap~\cite{memecap2023} or MET-Meme~\cite{10.1145/3477495.3532019} provide large-scale meme captioning resources, they lack cross-cultural pairs required for transcreation. Our work extends transcreation to memes, which require tight visual–textual coupling and humor preservation, and introduces a bidirectional benchmark spanning US and Chinese cultures.

\paragraph{Generative Vision–Language Models.}
Recent VLMs demonstrate strong multimodal understanding and reasoning capabilities, including LLaVA~\cite{liu2023visualinstruction, liu2024llavanext}, GPT-4V~\cite{10.1145/3729239}, and Qwen-VL~\cite{Qwen-VL, wang2024qwen2vl}. Parallel advances in image generation models enable increasingly faithful prompt-based visual synthesis~\cite{flux2023, flux1-lite}. While these models provide the foundation for multimodal generation, their effectiveness for culturally grounded creative adaptation remains underexplored—a gap our study aims to address.

\paragraph{Evaluation of Multimodal Generation.}
Standard metrics such as CLIPScore~\cite{hessel-etal-2021-clipscore} and TIFA~\cite{hu2023tifa} focus on text–image alignment but are not designed to capture cultural fit or humor preservation. Existing cross-cultural benchmarks, including CVQA~\cite{romero2024cvqaculturallydiversemultilingualvisual}, GlobalRG~\cite{bhatia-etal-2024-local}, and WorldCuisines~\cite{winata2025worldcuisines}, primarily address visual question answering rather than generative tasks~\cite{bai-etal-2025-power, Bhalerao2025MultiAgentMM}. In this work, we evaluate meme transcreation using both human judgments and automated LLM-based evaluation across multiple quality dimensions, enabling a systematic comparison of human and automated assessments in a cross-cultural setting.

\section{Hybrid Transcreation Framework}

We introduce a hybrid framework for cross-cultural meme transcreation that balances preservation of communicative intent with culturally appropriate adaptation. Rather than framing memes as a translation task, \emph{our approach explicitly separates culture-invariant elements from those that must change to ensure cultural authenticity}. This section outlines the guiding principles of the framework and the three-stage pipeline that implements them.

In practice, memes combine universal and culture-specific components: literal translation often preserves surface meaning but fails culturally, while full recreation risks drifting from the original intent. To address this trade-off, our hybrid strategy is grounded in three principles.

\noindent\textbf{Preserve universal elements.} We retain transferable aspects such as core humor mechanisms (e.g., irony, exaggeration), high-level emotional intent, and common meme formats.

\noindent\textbf{Adapt culture-specific elements.} We identify and replace culturally grounded references—such as pop culture, idioms, visual symbols, and stylistic conventions—with culturally appropriate alternatives rather than literal translations.

\noindent\textbf{Maintain intent consistency.} Across all stages, we preserve the meme's communicative goal—\textit{what it aims to express or satirize}—even when textual and visual inputs change.

\subsection{Meme Transcreation Pipeline}

\begin{figure}[t]
\centering
\includegraphics[width=\columnwidth]{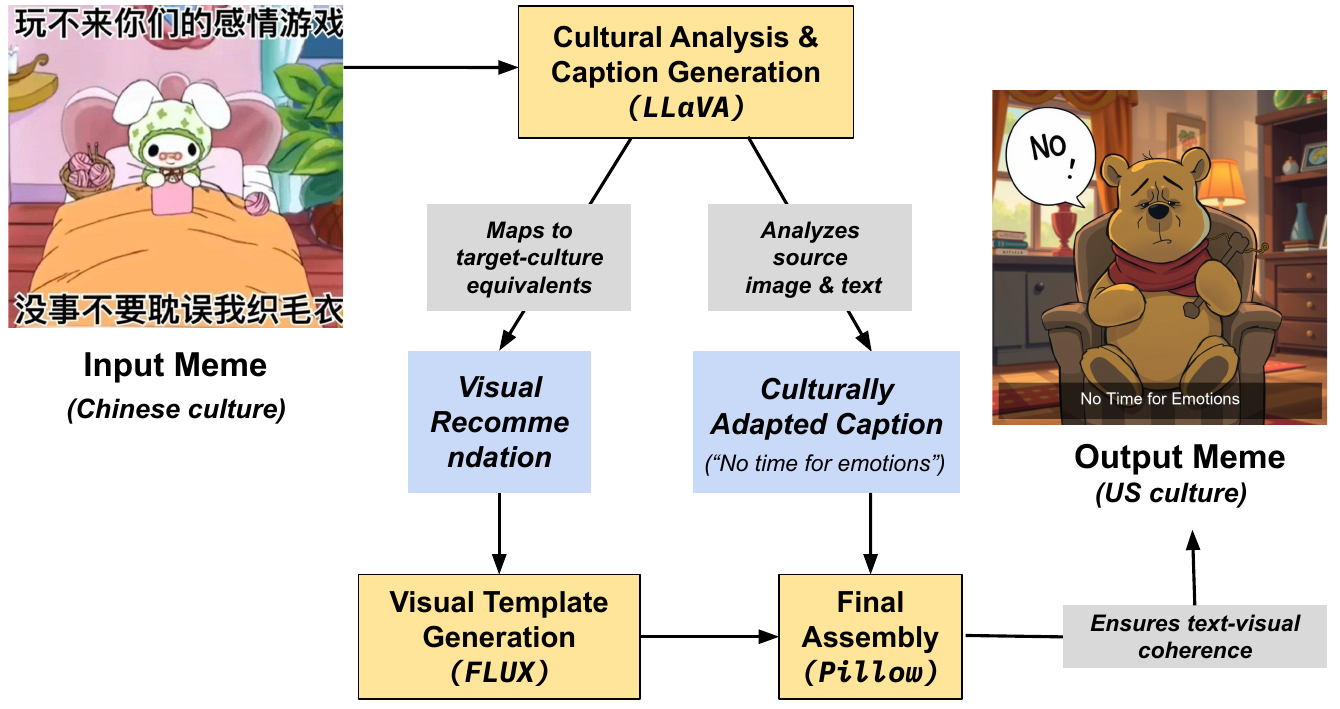}
\caption{Overview of our three-stage meme transcreation pipeline. (1) A VLM analyzes the \textit{original/input} meme, identifies cultural references and intent, and generates a culturally adapted caption. (2) A diffusion model produces a meme-style visual template aligned with the target culture. (3) A text overlay module assembles the output \textit{transcreated} meme.}
\label{fig:pipeline}
\vspace{-1.5em}
\end{figure}

Figure~\ref{fig:pipeline} illustrates our modular three-stage meme transcreation pipeline: \emph{cultural reasoning}, \emph{visual generation}, and \emph{final assembly}, enabling independent control and analysis across cultural directions.

\noindent\textbf{Stage 1: Cultural Analysis and Caption Generation.}
We use LLaVA~1.6 (13B)~\cite{liu2024llavanext} as the core vision-language model for cultural analysis and caption generation. The model takes the original meme image as input and is prompted to (1) analyze cultural references and humor mechanisms, (2) extract the underlying intent, (3) map source-culture concepts to target-culture equivalents, and (4) generate a culturally appropriate caption in the target language. 

We employ LLaVA~1.6 because it offers strong vision--language alignment, robust Chinese and English multilingual performance, and open-source reproducibility. Importantly, we do not fine-tune the model, focusing instead on prompt-based control to isolate the effects of cultural reasoning without introducing task-specific training bias.  This stage outputs both a transcreated caption and high-level recommendations for adapting the visual content (i.e., mood, background - examples in \Cref{sec:pipeline})

\noindent\textbf{Stage 2: Visual Template Generation.}
Using the visual recommendations from Stage~1, we generate culturally adapted meme templates with FLUX.1 Schnell~\cite{flux2023,flux1-lite}, a diffusion-based image generation model designed for strong prompt adherence and fast iteration. At this stage, the goal is not photorealism but meme-appropriate visuals that support the intended humor and allow for clear and readable text overlay.
The generated images preserve universal meme structures (e.g., reaction shots, simple backgrounds) while adapting culture-specific elements. For example, Western celebrity figures are often replaced with symbolic or animal-based representations that are more common in Chinese meme culture. Emotional tone is conveyed through facial expressions, posture, and visual metaphors that align with conventions in the target culture.

\noindent\textbf{Stage 3: Final Assembly.}
In the final stage, we automatically combine transcreated captions with the generated visual templates using Pillow, an open-source image processing library.\footnote{\url{https://python-pillow.org/}} This step handles font selection, text placement, and layout conventions appropriate for the target culture (e.g., denser layouts for Chinese text and more spaced layouts for English captions), following common social media meme practices.
We apply dynamic text wrapping, semi-transparent background overlays for readability, and center-aligned multi-line captions positioned near the image bottom. Final manual quality checks verify readability, visual--text coherence, and that captions do not obscure key visual elements.

\noindent\textbf{Implementation Details.}
All models are used in their pre-trained form, with prompt engineering (Appendix~\ref{sec:prompts}) and decoding parameter tuning (e.g., temperature, top-$p$) to balance creativity and consistency. This modular design supports reproducibility, scalability across cultures, and controlled analysis of cross-cultural meme transcreation.

\section{MemeXGen Dataset}
\label{sec:dataset}

To study cross-cultural meme transcreation in a controlled and realistic setting, we introduce \textbf{MemeXGen}, a multilingual and multicultural dataset of Chinese and US\ meme pairs. The dataset consists of 6,315 \textit{original memes} collected from authentic social media platforms and 6,315 \textit{transcreated memes} produced by our pipeline.
For each original meme, we generate a corresponding transcreated version in the opposite cultural context, resulting in a total of \textit{6,315 bidirectional meme pairs: 3{,}165 Chinese$\rightarrow$US\ and 3{,}150 US$\rightarrow$Chinese}. This paired structure enables direct comparison of transcreation quality across directions.



\subsection{Data Sources}
MemeXGen is designed to support systematic evaluation and analysis, with an emphasis on cultural authenticity, diversity of humor styles, and balanced coverage across Chinese and US cultures.

\noindent\textbf{Chinese Memes.}
\textit{Original} Chinese memes are sourced from the publicly available \textit{Chinese Meme Description Dataset}\footnote{\url{https://github.com/THUDM/chinese-meme-description-dataset}}, which aggregates content from two major Chinese social media platforms: \textbf{Xiaohongshu} and \textbf{Weibo}. Xiaohongshu contributes lifestyle- and emotion-focused memes, while Weibo provides fast-paced, commentary-driven content reflecting mainstream Chinese internet culture.  

\noindent\textbf{US Memes.}
\textit{Original} US memes are drawn from the \textit{MemeCap} dataset\footnote{\url{https://github.com/hwang1996/MemeCap}}, which collects memes from popular Reddit communities such as \texttt{r/memes} and \texttt{r/dankmemes}. These memes reflect dominant US humor styles, including sarcasm, irony, pop culture references, and situational storytelling.  

These sources offer complementary views of meme culture in two distinct cultural contexts, enabling systematic bidirectional transcreation.

\subsection{Filtering and Dataset Composition}
During data inspection, we observe that some \textit{original} memes contain potentially sensitive content (e.g., political references) that could interfere with fair evaluation or raise ethical concerns. To address this, we manually filter the \textit{original} memes to ensure responsible use and reliable evaluation.
Specifically, we remove memes that are offensive, contain low-quality or corrupted images, rely heavily on mixed languages, or exhibit weak visual--text integration. After filtering, the dataset contains 6{,}315 \textit{original} memes, equally split across US and Chinese.
We notice that the retention rate is substantially higher for the Chinese subset (97.6\%) than for the US\ subset (55.0\%), reflecting stricter content moderation on Chinese platforms compared to the more permissive nature of Reddit.

\subsection{Annotation and Evaluation Split}

To support emotion analysis and human evaluation, we annotate two disjoint subsets of the data.  

\noindent\textbf{Emotion annotations subset.} We annotate $\approx$10\% of the \textit{original} memes data (628 memes, equally split between US\ and Chinese memes) with emotion labels, including emotion category (\textit{Joy, Anger, Sadness, Fear, Disgust, Surprise}) and emotion intensity on a 1--5 Likert scale. Annotation guidelines follow recent multilingual emotion classification work, as described in BRIGHTER~\cite{muhammad-etal-2025-brighter}. Three expert annotators perform the annotations independently, achieving strong agreement (Fleiss’~\cite{Fleiss1971} $\kappa = 0.869$ for emotion category and $\kappa = 0.536$ for intensity).

\noindent\textbf{Human evaluation subset.} We reserve a separate 10\% subset (628 \textit{original} memes) as the test set for transcreation experiments. This split includes 313 Chinese$\rightarrow$US and 315 US$\rightarrow$Chinese \textit{original}-\textit{transcreated} meme pairs and is used exclusively for human evaluation of meme transcreation. Evaluation details are provided in Section \ref{sec:evaluation}.

\subsection{Dataset Characteristics}
To better understand the cultural makeup of the \textit{original} memes, we analyze topic and emotion distributions using Qwen-VL-Max~\cite{Qwen-VL,wang2024qwen2vl}, finetuned on the human annotated emotions. 
To validate the reliability of the predicted labels, an expert annotator manually reviews a random 10\% subset and confirms that over 95\% of the topic and emotion labels are correct.

\noindent\textbf{Topic Distribution.}
Both cultures are dominated by themes related to \textit{Internet Culture} (CN 61.0\%, US\ 52.4\%) and \textit{Technology/Digital Life} (CN 10.6\%, US\ 15.1\%). Beyond these shared themes, clear differences emerge. \textbf{US\ memes more often frame education, family, and everyday experiences as relatable, narrative-driven humor} (e.g., \textit{Education}: 7.8\%; \textit{Family}: 4.9\%), whereas \textbf{Chinese memes emphasize symbolic expression and social pressure}, with lower prevalence of these topics (\textit{Education}: 2.1\%; \textit{Family}: 1.9\%). \textit{Gaming}-related humor appears among the top US\ topics (2.7\%) but is largely absent from the Chinese top ranks, reflecting differing leisure and achievement orientations.

\noindent\textbf{Emotion Distribution.}
Automated emotion classification shows that \textbf{\textit{Joy} dominates in both cultures} (CN 69.3\%, US\ 73.8\%), consistent with memes’ primary entertainment role. However, Chinese memes exhibit higher levels of \textit{Anger} (8.3\%) and \textit{Sadness} (8.2\%), suggesting more frequent \textbf{social critique and melancholic expression}. In contrast, US\ memes show relatively higher \textit{Fear} (7.0\%) and \textit{Disgust} (4.4\%), aligning with \textbf{anxiety-driven and cringe-based humor styles}.

These systematic differences motivate our hybrid transcreation approach and provide context for interpreting performance asymmetries in later experiments. Further data analysis is provided in \Cref{sec:data}.

\section{Evaluation Methodology}
\label{sec:evaluation}
We evaluate our meme transcreation framework using both human and VLM-based evaluation.

\subsection{Metric Definitions}

Our evaluation captures not only text and image quality and their interaction, as commonly assessed in prior image generation work~\cite{hu2023tifa}, but also \textit{task-specific} aspects that are critical for cross-cultural transcreation, namely cultural fit and intent preservation. All quantitative metrics are rated on a 5-point Likert scale (1 = strongly disagree, 5 = strongly agree).

We evaluate each transcreated meme along six dimensions: \textbf{Caption Quality}, measuring clarity, tone, and meme-appropriate language; \textbf{Image Quality}, assessing visual clarity, composition, and recognizability; \textbf{Synergy}, capturing how well image and text work together to convey humor or emotion; \textbf{Cultural Fit}, evaluating appropriateness and relatability for the target culture; \textbf{Intent Preservation}, measuring retention of the original meme’s message and emotional effect; and an \textbf{Overall Score}, computed as the average across all dimensions. Detailed dimension definitions are provided in \Cref{sec:metrics}.


\subsection{Human Evaluation}
We evaluate meme transcreation on the human evaluation subset, with each meme independently assessed by three human evaluators. Because meme generation is inherently subjective and prior work highlights the importance of modeling annotator perspectives rather than simple aggregation~\cite{deng-etal-2023-annotate}, we report results separately for each evaluator. All evaluators are bilingual and bicultural, with deep familiarity with both Chinese and US\ meme cultures. Additional details on annotator backgrounds are provided in \Cref{sec:evaluators}.

\noindent\textbf{Inter-annotator Agreement.}
Inter-annotator Pearson correlations indicate moderate to strong agreement ($r = 0.58$--$0.81$), reflecting reliable yet stylistically distinct evaluation perspectives.

\subsection{Automatic Evaluation}
To assess the feasibility of automated evaluation, we use six state-of-the-art VLMs on all the data: Qwen-VL-Max~\cite{Qwen-VL}, LLaVA-v1.6-Vicuna-13B~\cite{liu2024llavanext}, InternVL3-8B and InternVL3-14B~\cite{zhu2025internvl3}, Qwen3-VL-8B~\cite{wang2024qwen2vl}, and Aya-vision-8b~\cite{dash2025ayavision}. These models are selected for their multilingual Chinese--English support, ability to process multiple images, and public availability, enabling reproducible automatic evaluation.

\noindent\textbf{VLM-Human Agreement.} We asses VLM evaluation effectiveness by computing the Pearson correlation between each human evaluator and VLM across each dimension (results in \Cref{sec:eval}).

\begin{table*}[t!]
\centering
\footnotesize
\setlength{\tabcolsep}{5pt}
\renewcommand{\arraystretch}{1.15}
\begin{tabular}{lcccccc|cccccc}
\toprule
\multicolumn{7}{c|}{\textbf{Chinese$\rightarrow$US}} &
\multicolumn{6}{c}{\textbf{US$\rightarrow$Chinese}} \\
\midrule
\textbf{Evaluator} &
\textbf{Cap.} & \textbf{Img.} & \textbf{Syn.} & \textbf{Cult.} & \textbf{Intent} & \textbf{Overall} &
\textbf{Cap.} & \textbf{Img.} & \textbf{Syn.} & \textbf{Cult.} & \textbf{Intent} & \textbf{Overall} \\
\midrule
Evaluator 1 (H) &
4.78 & 4.51 & 4.66 & 4.57 & 4.24 & 4.55 &
4.82 & 4.22 & 4.31 & 4.18 & 3.89 & 4.28 \\
Evaluator 2 (H) &
4.35 & 4.11 & 4.45 & 4.14 & 4.00 & 4.21 &
4.41 & 3.84 & 4.12 & 3.76 & 3.67 & 3.96 \\
Evaluator 3 (H) &
3.46 & 3.52 & 3.59 & 3.39 & 3.29 & 3.45 &
3.52 & 3.19 & 3.24 & 2.98 & 2.93 & 3.17 \\
\midrule
Qwen-VL-Max     &
\textbf{4.13} & 3.86 & \textbf{4.20} & \textbf{3.74} & 3.72 & \textbf{3.95} &
\textbf{4.21} & 3.58 & \textbf{3.89} & \textbf{3.41} & 3.44 & \textbf{3.71} \\
LLaVA-v1.6      &
\underline{4.00} & \underline{4.00} & \underline{3.81} & 3.00 & \textbf{4.00} & \underline{3.79} &
\underline{4.05} & 3.72 & \underline{3.52} & 2.67 & \textbf{3.71} & \underline{3.53} \\
InternVL3-8B    &
3.78 & 3.84 & 3.48 & 3.46 & \underline{3.97} & 3.69 &
3.84 & 3.55 & 3.16 & 3.12 & \underline{3.68} & 3.47 \\
InternVL3-14B   &
3.21 & \textbf{4.39} & 3.16 & 3.36 & 3.34 & 3.53 &
3.28 & \textbf{4.11} & 2.84 & 3.02 & 3.01 & 3.25 \\
Qwen3-VL-8B     &
2.83 & 3.70 & 2.74 & \underline{3.59} & 2.56 & 3.15 &
2.91 & 3.42 & 2.41 & \underline{3.21} & 2.18 & 2.83 \\
Aya-vision-8b   &
3.18 & \underline{4.17} & 2.83 & 2.90 & 2.72 & 3.10 &
3.25 & \underline{3.89} & 2.51 & 2.56 & 2.39 & 2.92 \\
\bottomrule
\end{tabular}

\vspace{0.5em}
\footnotesize
\textit{Note:} Best VLM results per column are shown in \textbf{bold}, second-best VLM results are \underline{underlined}.  
All dimensions rated 1--5 (higher = better). (H) = Human evaluator.  
Cap.=Caption Quality, Img.=Image Quality, Syn.=Synergy, Cult.=Cultural Fit, Intent=Intent Preservation.
\caption{Evaluation Results for Chinese$\rightarrow$US and US$\rightarrow$Chinese Meme Transcreation}
\label{tab:main_results}
\end{table*}

\section{Evaluation Results}

We report evaluation results addressing our research questions, using human and automatic metrics to assess cross-cultural performance, directional effects, and evaluation reliability.

\subsection{RQ1: Cross-Cultural Performance}

\noindent\textbf{Human Evaluation.}
Table~\ref{tab:main_results} summarizes results from three human evaluators and six LLM evaluators across both transcreation directions. Human evaluators differ in strictness and focus: Evaluator~1 prioritizes entertainment value (mean: 4.42), Evaluator~2 adopts a balanced perspective (mean: 4.09), and Evaluator~3 applies stricter quality standards (mean: 3.31). The resulting 1.11-point spread highlights the inherent subjectivity of cross-cultural meme transcreation evaluation. \textbf{Overall, the mean human score of 4.07/5.0 indicates that the proposed pipeline produces effective and generally well-received transcreations.}

\noindent\textbf{Dimension-Level Analysis.}
Across dimensions, \textit{Caption Quality} receives the highest ratings (mean: 4.20), suggesting effective cross-cultural adaptation of meme text. \textit{Image Quality} is also strong (mean: 4.05), supporting the reliability of FLUX.1 for meme-style visual generation. \textbf{\textit{Synergy} is consistently high (mean: 4.23)}, indicating that captions and visuals work well together in most outputs. \textit{Cultural Fit} shows the widest variation across evaluators (range: 3.39--4.57), reflecting the subjective and culturally grounded nature of authenticity judgments. \textit{Intent Preservation} is rated favorably overall (mean: 3.84), though scores may be partially influenced by the dominance of Joy-related memes (69--74\% of the data). 

\noindent\textbf{VLM Evaluation.}
Among automated evaluators, \textbf{Qwen-VL-Max performs best}, achieving the highest overall scores (3.95 for Chinese$\rightarrow$US and 3.71 for US$\rightarrow$Chinese) and showing \textbf{strong alignment with human judgments} (mean Pearson $r = 0.926$, all $p < 0.001$). Other LLM evaluators exhibit much weaker correlations with humans ($r \leq 0.33$), suggesting that most open-source models struggle to reliably evaluate creative, culturally grounded outputs. \textbf{On average, LLM scores are 0.54 points lower than human scores}, indicating a systematic tendency toward conservative scoring.

\subsection{RQ2: Directional Asymmetries}

We observe a clear directional asymmetry: \textbf{US$\rightarrow$Chinese meme transcreations significantly outperform Chinese$\rightarrow$US} (4.48 vs.\ 3.93 out of 5.0, $\delta = +0.55$, $p < 0.001$). This gap likely reflects several factors. First, US\ memes often rely on globally recognizable characters and themes that are easier to localize, whereas Chinese memes frequently depend on context-specific wordplay and implicit cultural knowledge. Second, current VLMs are more strongly exposed to US-centric data during training. Third, evaluators apply stricter authenticity expectations to Chinese$\rightarrow$US outputs, where cultural mismatches are more salient to native speakers.

\subsection{RQ3: Evaluation Framework Analysis}
\label{sec:eval}

Table~\ref{tab:correlations} reports correlations between human and LLM evaluators. \textbf{Qwen-VL-Max shows consistently strong alignment with all three human evaluators} (Evaluator~1: $r = 0.921$, Evaluator~2: $r = 0.964$, Evaluator~3: $r = 0.894$), with especially high agreement on \textit{Intent Preservation} ($r = 0.993$). In contrast, other models exhibit substantially weaker correlations (e.g., InternVL3-14B: $r = 0.263$, Qwen3-VL-8B: $r = 0.252$, LLaVA-v1.6: $r = 0.005$). \textbf{These results suggest that evaluating creative, cross-cultural multimodal content requires deeper multilingual and multicultural grounding than most current open-source VLMs provide.}

\begin{table}[t]
\centering
\footnotesize
\resizebox{\columnwidth}{!}{%
\begin{tabular}{llcccccc}
\toprule
\textbf{Human} & \textbf{VLM} & \textbf{Cap.} & \textbf{Img.} & \textbf{Syn.} & \textbf{Cult.} & \textbf{Intent} & \textbf{Overall} \\
\midrule
\multirow{6}{*}{Eval. 1} 
& Qwen-VL-Max & \corrcolor{0.961} & \corrcolor{0.986} & \corrcolor{0.987} & \corrcolor{0.901} & \corrcolor{0.993} & \corrcolor{0.921} \\
& LLaVA-v1.6 & \corrcolor{-0.039} & \corrcolor{-0.039} & \corrcolor{0.082} & \corrcolor{-0.178} & \corrcolor{-0.039} & \corrcolor{-0.026} \\
& InternVL3-8B & \corrcolor{-0.128} & \corrcolor{-0.026} & \corrcolor{-0.053} & \corrcolor{-0.106} & \corrcolor{0.041} & \corrcolor{-0.086} \\
& InternVL3-14B & \corrcolor{0.241} & \corrcolor{0.363} & \corrcolor{0.288} & \corrcolor{0.216} & \corrcolor{0.275} & \corrcolor{0.270} \\
& Qwen3-VL-8B & \corrcolor{0.219} & \corrcolor{0.394} & \corrcolor{0.316} & \corrcolor{0.215} & \corrcolor{0.289} & \corrcolor{0.281} \\
& Aya-vision-8b & \corrcolor{-0.117} & \corrcolor{-0.032} & \corrcolor{-0.087} & \corrcolor{-0.077} & \corrcolor{-0.065} & \corrcolor{-0.082} \\
\midrule
\multirow{6}{*}{Eval. 2} 
& Qwen-VL-Max & \corrcolor{0.989} & \corrcolor{0.988} & \corrcolor{0.994} & \corrcolor{0.980} & \corrcolor{0.995} & \corrcolor{0.964} \\
& LLaVA-v1.6 & \corrcolor{-0.016} & \corrcolor{-0.016} & \corrcolor{0.038} & \corrcolor{-0.112} & \corrcolor{-0.016} & \corrcolor{-0.027} \\
& InternVL3-8B & \corrcolor{-0.099} & \corrcolor{0.014} & \corrcolor{-0.033} & \corrcolor{-0.061} & \corrcolor{0.041} & \corrcolor{-0.066} \\
& InternVL3-14B & \corrcolor{0.294} & \corrcolor{0.350} & \corrcolor{0.351} & \corrcolor{0.323} & \corrcolor{0.365} & \corrcolor{0.329} \\
& Qwen3-VL-8B & \corrcolor{0.265} & \corrcolor{0.340} & \corrcolor{0.331} & \corrcolor{0.294} & \corrcolor{0.327} & \corrcolor{0.309} \\
& Aya-vision-8b & \corrcolor{-0.088} & \corrcolor{0.022} & \corrcolor{-0.048} & \corrcolor{-0.034} & \corrcolor{-0.031} & \corrcolor{-0.058} \\
\midrule
\multirow{6}{*}{Eval. 3} 
& Qwen-VL-Max & \corrcolor{0.950} & \corrcolor{0.990} & \corrcolor{0.985} & \corrcolor{0.938} & \corrcolor{0.990} & \corrcolor{0.894} \\
& LLaVA-v1.6 & \corrcolor{-0.050} & \corrcolor{-0.050} & \corrcolor{0.039} & \corrcolor{-0.161} & \corrcolor{-0.050} & \corrcolor{0.029} \\
& InternVL3-8B & \corrcolor{-0.065} & \corrcolor{0.083} & \corrcolor{0.014} & \corrcolor{0.008} & \corrcolor{0.107} & \corrcolor{-0.013} \\
& InternVL3-14B & \corrcolor{0.296} & \corrcolor{0.348} & \corrcolor{0.373} & \corrcolor{0.320} & \corrcolor{0.371} & \corrcolor{0.340} \\
& Qwen3-VL-8B & \corrcolor{0.193} & \corrcolor{0.368} & \corrcolor{0.310} & \corrcolor{0.235} & \corrcolor{0.285} & \corrcolor{0.263} \\
& Aya-vision-8b & \corrcolor{-0.024} & \corrcolor{0.101} & \corrcolor{0.016} & \corrcolor{0.050} & \corrcolor{0.025} & \corrcolor{-0.057} \\
\bottomrule
\end{tabular}
}
\vspace{0.5em}
\footnotesize
\textit{Note:}  Cap.=Caption Quality, Img.=Image Quality, Syn.=Synergy, Cult.=Cultural Fit, Intent=Intent Preservation.
Cell colors indicate correlation strength.
\caption{Pearson correlation coefficients ($r$) between Human and LLM evaluators across evaluation dimensions.}

\label{tab:correlations}
\end{table}

\subsection{Qualitative Analysis}

\begin{figure*}[t]
\centering
\begin{subfigure}[t]{0.48\textwidth}
    \centering
    \includegraphics[width=\linewidth]{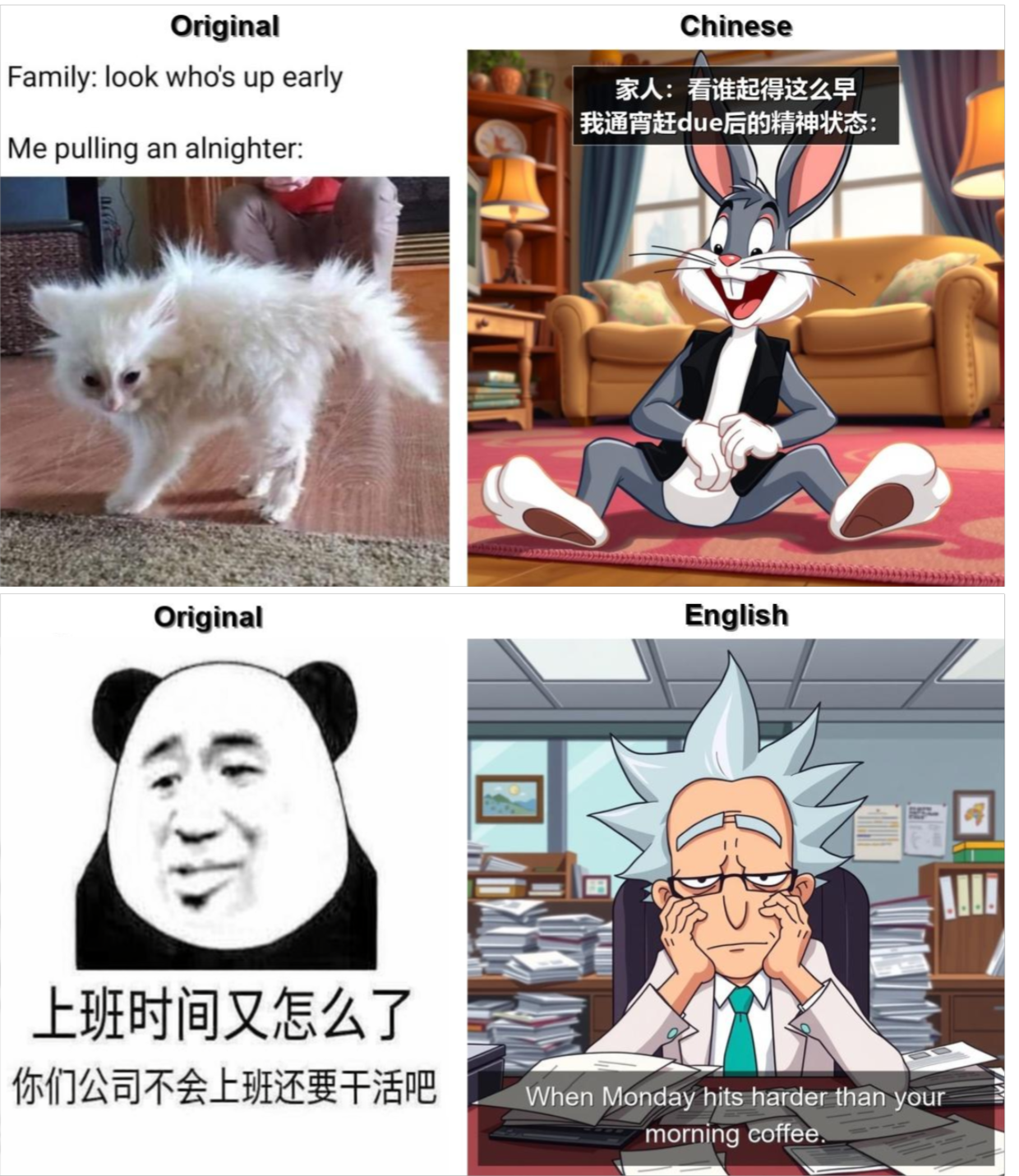}
    \caption{
    \textbf{Best transcreation example} (\textbf{Score: 5.0/5.0}).\\
    \textbf{\textcolor{emphblue}{Transcreated (Chinese, top right — Bugs Bunny):}} 
    ``Family: Look who's up so early; My mental state after pulling an all-nighter to finish my assignment due.''\\
    \textbf{\textcolor{emphblue}{Original (bottom left — panda meme):}} 
    ``What's wrong with work hours? Doesn't your company expect you to work during work hours?''
    }
    \label{fig:qual_en_cn}
\end{subfigure}
\hfill
\begin{subfigure}[t]{0.48\textwidth}
    \centering
    \includegraphics[width=\linewidth]{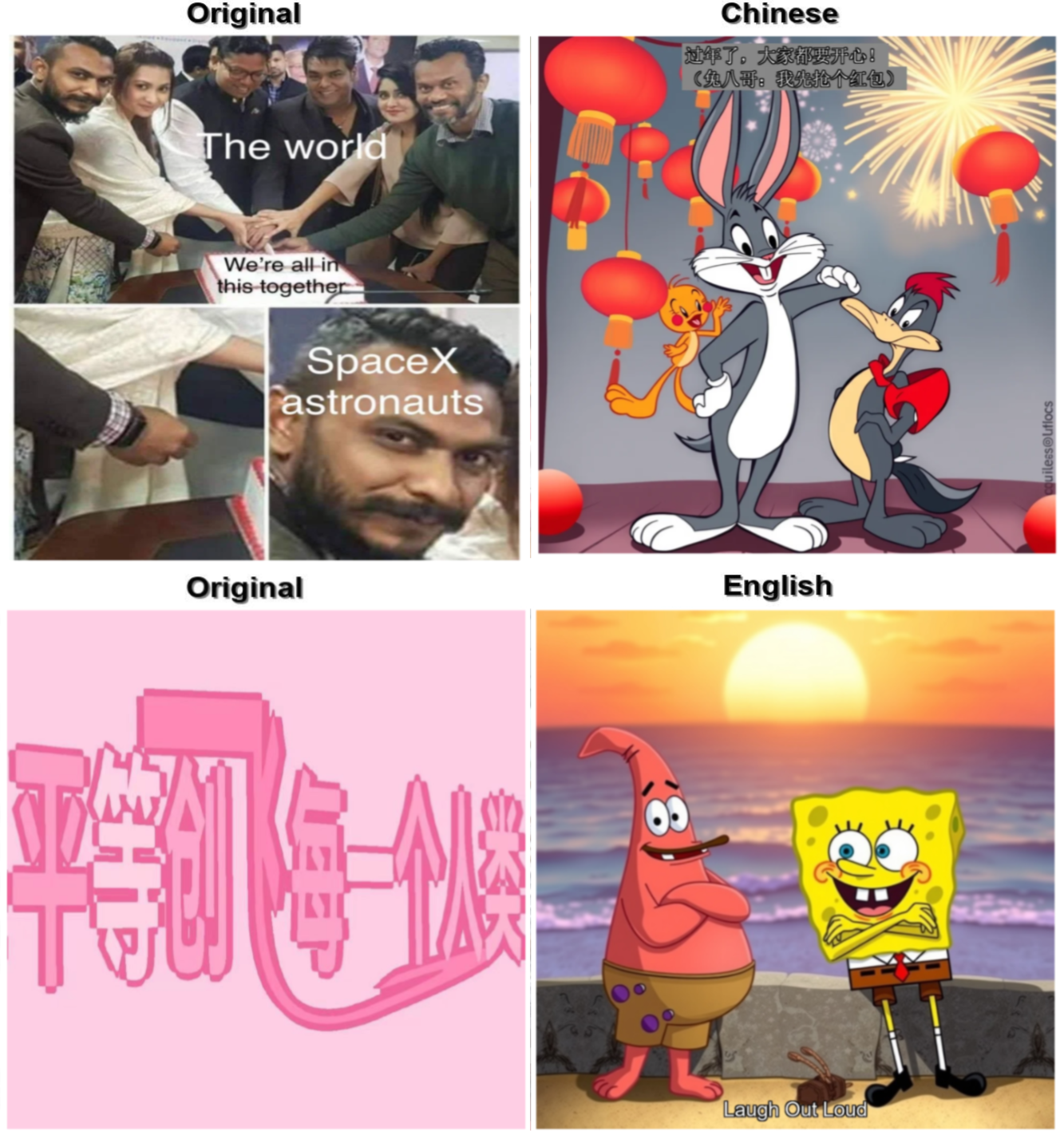}
    \caption{
    \textbf{Worst transcreation example} (\textbf{Score: 1.4/5.0}).\\
    \textbf{\textcolor{emphred}{Transcreated (Chinese, top right — Looney Tunes):}} 
    ``Celebrating the New Year, everyone must be happy! No one left behind, I'm giving you a family photo!''\\
    \textbf{\textcolor{emphred}{Original (bottom left — Chinese text):}} 
    ``Attempts at equality satisfy no one.''
    }
    \label{fig:qual_cn_en}
\end{subfigure}

\caption{
\textbf{Qualitative examples of cross-cultural meme transcreation.}
\textcolor{emphblue}{Left:} successful US$\rightarrow$Chinese adaptation preserving intent, humor, and cultural conventions.
\textcolor{emphred}{Right:} failed Chinese$\rightarrow$US adaptation illustrating loss of intent and cultural mismatch.
}
\label{fig:qualitative_examples}
\end{figure*}

To complement quantitative metrics, we present representative transcreation examples from both directions and our data analysis observations. Figure~\ref{fig:qual_en_cn} shows two successful transcreation samples from both directions (Overall Score: 5.0/5.0), while Figure~\ref{fig:qual_cn_en} illustrates several failed transcreation samples (Overall Score: 1.4/5.0).

\noindent\textbf{Success Patterns (30\% of outputs scoring $\geq$4.5/5.0).} High quality meme transcreations contain the following elements: (1) \textit{Universally applicable character selection}—the use of recognizable archetypes that can be understood across cultures. (2) \textit{Emotion-focused transcreations}—the retention of the original emotional context with the incorporation of cultural specifics. (3) \textit{Use of natural language conventions}—the use of meme-like linguistic conventions associated with the receiving culture. (4) \textit{Visual and textual unity}—careful matching of image and text

\noindent\textbf{Failure Patterns (1.6\% of outputs scoring $\leq$2.0/5.0).} Errors that emerge on failed meme transcreations include: (1) \textit{Failure in Captions}—the use of formal speech that dampens the casual meme vibe; (2) \textit{Disconnects in Visuals}—the use of images that don't fit a culture or issues with visual generation; (3) \textit{Failure to Preserve Humor Mechanisms}—the use of a format that is not amenable to joke structure; (4) \textit{Complete synergy breakdown}—caption-image mismatch creating incoherent messaging.

\noindent\textbf{Directional Patterns.} US$\rightarrow$Chinese transcreations more frequently achieve natural cultural integration, benefiting from globally recognizable US templates. Chinese$\rightarrow$US transcreations struggle with context-dependent wordplay and philosophical concepts that lack Western equivalents, often resulting in superficial adaptations. Additional examples in Appendix~\ref{sec:examples}.

\section{Main Takeaways}
\label{sec:discussion}

\noindent\textbf{Effective Cross-Cultural Transcreation.}
Human evaluations show that the proposed hybrid approach produces high-quality transcreations (mean score: 4.07/5.0), successfully preserving humor and intent while adapting cultural specifics. Strong performance on Caption Quality (4.20) and Image--Text Synergy (4.23) confirms that the three-stage pipeline supports coherent multimodal generation.

\noindent\textbf{Directional Asymmetry Matters.}
US$\rightarrow$Chinese transcreation consistently outperforms Chinese$\rightarrow$US (+0.55), reflecting both model exposure biases and deeper cultural differences. In particular, Chinese memes rely more heavily on context-dependent wordplay and implicit meaning, which are harder to adapt than the more visually universal templates common in US meme culture. These results highlight the need for culturally diverse training data in cross-cultural AI systems.

\noindent\textbf{Limits of Automated Evaluation.}
Qwen-VL-Max shows strong agreement with human judgments ($r = 0.926$), demonstrating that automated evaluation of creative, cross-cultural content is feasible. However, weaker correlations from open-source models suggest that reliable automated evaluation remains challenging without extensive multilingual and multicultural grounding.

\section{Conclusion}
\label{sec:conclusion}

We introduced a hybrid framework for cross-cultural meme transcreation that explicitly separates intent preservation from cultural adaptation, enabling principled analysis of how humor and meaning transfer across cultures. By combining vision--language models with diffusion-based image generation, our approach moves beyond literal translation and treats meme adaptation as a culturally grounded multimodal reasoning problem.

We curated and evaluated a dataset of 6{,}315 Chinese--U.S.\ meme pairs, combining authentic social media memes with systematically generated transcreations, and conducted a comprehensive bidirectional evaluation. Our results reveal consistent directional asymmetries in transcreation quality, demonstrating that current models handle certain cultural adaptations more effectively than others. These findings expose concrete limitations in cross-cultural generalization that are not visible in monolingual or translation-based evaluations.

We further show that carefully selected VLM-based evaluators can approximate human judgments on culturally grounded dimensions such as emotion and intent, while most open-source models remain unreliable for assessing intent and cultural fit. Finally, we release \textsc{MemeXGen}, the first parallel Chinese--U.S.\ meme transcreation corpus annotated for emotion and cultural intent, together with evaluation protocols and dataset splits. By open-sourcing data, models, and evaluation metrics, this work establishes a foundation for systematic study of computational humor and cross-cultural multimodal generation, and provides actionable benchmarks for future model development.





\section*{Limitations}

\paragraph{Scope of Cultural Coverage.}
This study focuses on meme transcreation between Chinese and U.S.\ cultures, which differ substantially in language, humor conventions, and visual symbolism. While this contrast enables clear analysis of cultural asymmetries, our findings should not be assumed to generalize uniformly to other cultural pairs. Future work should extend this framework to additional language–culture settings to test the robustness of the observed patterns.

\paragraph{Generality of the Generation Framework.}
Our transcreation pipeline combines existing vision--language and diffusion models in a modular design intended to support interpretability and controlled analysis rather than architectural novelty. We do not claim optimality of this design, nor do we compare against all possible end-to-end prompting alternatives. Instead, our goal is to provide a transparent framework for studying cultural adaptation. Exploring simpler or fully integrated baselines remains an important direction for future work.

\paragraph{Interpreting Directional Asymmetries.}
We observe consistent performance differences between US$\rightarrow$Chinese and Chinese$\rightarrow$US transcreation. While we discuss plausible contributing factors—such as training data exposure, humor structure, and evaluator expectations—these explanations are correlational rather than causal. Disentangling these effects would require controlled experiments that vary data distributions, model pretraining, and evaluation populations independently.

\paragraph{Limits of Automatic Evaluation.}
Although Qwen-VL-Max shows strong alignment with human judgments in our setting, this result may reflect model-specific strengths in Chinese--English multimodal understanding rather than a general solution to evaluating culturally grounded humor. The weak performance of other open-source evaluators highlights that reliable automated evaluation remains challenging and should be interpreted with caution.

\paragraph{Dataset Composition and Emotion Coverage.}
Joy dominates the meme distributions in both cultures, reflecting real-world social media trends but limiting stress-testing on negative or socially critical humor. As a result, intent preservation scores may be optimistic for emotionally complex cases. Expanding emotion-balanced datasets is a key area for future research.

\noindent\textbf{Evaluation at Scale and in the Wild.}
Human evaluation remains inherently subjective, and the observed variation across evaluators highlights the value of modeling diverse perspectives rather than collapsing them into a single score. Expanding the evaluation set and incorporating longitudinal, in-the-wild measurements (e.g., engagement or sharing behavior) would provide deeper insight into real-world cultural impact beyond offline quality ratings.

\noindent\textbf{Broadening Cultural Perspectives.}
Our annotators are bilingual and bicultural with Chinese--U.S.\ experience, ensuring informed evaluation of both contexts. Future work can further broaden cultural representation by including evaluators with more localized or region-specific backgrounds, as well as exploring regional variation within Chinese and U.S.\ meme cultures. Such diversity would deepen understanding of cultural nuance and strengthen the generalizability of cross-cultural evaluation.




\section*{Ethical Considerations}
\label{sec:ethics}

\paragraph{Deployment Scope.}
Our pipeline prioritizes analytical clarity and controlled study of cultural adaptation rather than deployment efficiency. As with any automated cultural generation system, misuse, misinterpretation, or oversimplification of cultural signals remains a risk. We position meme transcreation as a \emph{decision-support tool} intended to assist human creators and analysts, not as a fully autonomous content generator, and strongly recommend human oversight in real-world or sensitive deployments.

\noindent\textbf{Content Safety.}
We apply stringent manual filtering to exclude offensive or sensitive content, including hate speech, discriminatory media, explicit material, and political content. High \emph{Not Offensive} ratings (92.8\% from human evaluators and 96.6\% from LLM-based assessment) indicate the effectiveness of these safeguards. However, cultural sensitivity is inherently subjective: content acceptable in one cultural context may still be perceived as offensive in another. Our system therefore cannot guarantee zero harmful outputs and should be used with caution, particularly in public-facing applications.

\noindent\textbf{Cultural Respect and Representation.}
Automated cultural adaptation risks reinforcing stereotypes or reducing complex cultural practices to surface-level substitutions. While our hybrid framework explicitly separates intent preservation from cultural adaptation, evaluator feedback reveals occasional cases of shallow cultural mapping (e.g., direct visual substitution without deeper contextual grounding). These limitations highlight the importance of human-in-the-loop workflows, where automated transcreation outputs are treated as drafts rather than finalized content.

\noindent\textbf{Data Privacy and Attribution.}
All source memes are collected from publicly accessible platforms (Xiaohongshu and Weibo for Chinese memes; Reddit for U.S.\ memes) and do not contain personal identifying information. We respect the implicit consent associated with public content sharing, though proper attribution remains challenging for viral meme formats with unclear authorship. The dataset is intended strictly for research purposes, and we encourage responsible use consistent with platform norms and community standards.


\noindent\textbf{Misinformation Potential.} Meme transcreation tools could be misused to spread culturally-adapted misinformation or propaganda. We emphasize responsible deployment with content verification protocols and transparency about automated generation.

\bibliography{custom}

@misc{memecap2023,
  title={MemeCap: A Dataset for Captioning and Interpreting Memes},
  author={Hwang, EunJeong and Shwartz, Vered},
  year={2023},
  eprint={2305.13703},
  archivePrefix={arXiv},
  url={https://arxiv.org/abs/2305.13703}
}

@misc{liu2024llavanext,
    title={LLaVA-NeXT: Improved reasoning, OCR, and world knowledge},
    url={https://llava-vl.github.io/blog/2024-01-30-llava-next/},
    author={Liu, Haotian and Li, Chunyuan and Li, Yuheng and Li, Bo and Zhang, Yuanhan and Shen, Sheng and Lee, Yong Jae},
    month={January},
    year={2024}
}

@article{Qwen-VL,
  title={Qwen-VL: A Versatile Vision-Language Model for Understanding, Localization, Text Reading, and Beyond},
  author={Bai, Jinze and Bai, Shuai and Yang, Shusheng and Wang, Shijie and Tan, Sinan and Wang, Peng and Lin, Junyang and Zhou, Chang and Zhou, Jingren},
  journal={arXiv preprint arXiv:2308.12966},
  year={2023}
}

@misc{flux2023,
  author={{Black Forest Labs}},
  title={FLUX: Advanced Text-to-Image Generation Model},
  year={2024},
  howpublished={\url{https://github.com/black-forest-labs/flux}},
  note={GitHub Repository}
}

@misc{flux1-lite,
  title={Flux.1 Lite: Distilling Flux1.dev for Efficient Text-to-Image Generation},
  author={Verdú, Daniel and Martín, Javier},
  year={2024},
  howpublished={Hugging Face Model Hub},
  url={https://huggingface.co/Freepik/flux.1-lite-8B}
}

@article{nissenbaum2018internet,
  title={Internet memes as contested cultural capital: The case of 4chan's/b/board},
  author={Nissenbaum, Asaf and Shifman, Limor},
  journal={New Media \& Society},
  volume={19},
  number={4},
  pages={483--501},
  year={2018},
  url={https://journals.sagepub.com/doi/10.1177/1461444817712511}
}

@inproceedings{hershcovich-etal-2022-challenges,
  title = "Challenges and Strategies in Cross-Cultural {NLP}",
  author = "Hershcovich, Daniel and others",
  booktitle = "Proceedings of the 60th Annual Meeting of the ACL",
  year = "2022",
  pages = "6997--7013",
  url={https://aclanthology.org/2022.acl-long.482/}
}

@inproceedings{field_survey_2021,
  title = {A {Survey} of {Race}, {Racism}, and {Anti}-{Racism} in {NLP}},
  booktitle = {Proceedings of the 59th {Annual} {Meeting} of the {ACL}},
  author = {Field, Anjalie and Blodgett, Su Lin and Waseem, Zeerak and Tsvetkov, Yulia},
  year = {2021},
  pages = {1905--1925},
  url={https://aclanthology.org/2021.acl-long.149/}
}

@misc{khanuja2024image,
  title={An image speaks a thousand words, but can everyone listen? On image transcreation for cultural relevance},
  author={Khanuja, Simran and Ramamoorthy, Sathyanarayanan and Song, Yueqi and Neubig, Graham},
  year={2024},
  eprint={2404.01247},
  archivePrefix={arXiv},
  url={https://arxiv.org/abs/2404.01247}
}

@misc{zhu2025internvl3,
  title={InternVL3: Exploring Advanced Training and Test-Time Recipes for Open-Source Multimodal Models},
  author={Zhu, Jinguo and Wang, Weiyun and Chen, Zhe and Liu, Zhaoyang and Ye, Shenglong and Gu, Lixin and Tian, Hao and Duan, Yuchen and Su, Weijie and Shao, Jie and others},
  year={2025},
  eprint={2504.10479},
  archivePrefix={arXiv},
  url={https://arxiv.org/abs/2504.10479}
}

@article{Fleiss1971,
  author = {Fleiss, Joseph L.},
  title = {Measuring nominal scale agreement among many raters},
  journal = {Psychological Bulletin},
  volume = {76},
  number = {5},
  pages = {378--382},
  year = {1971},
  doi = {10.1037/h0031619}
}

@misc{dash2025ayavision,
  title={Aya Vision: Advancing the Frontier of Multilingual Multimodality},
  author={Dash, Saurabh and Singh, Shivalika and Morisot, Adrien and Ermis, Beyza and Locatelli, Acyr and Hong, Sungjin and Ahmadian, Arash and Flet-Berliac, Yannis and Grinsztajn, Nathan and Strub, Florian and others},
  year={2025},
  eprint={2505.08751},
  archivePrefix={arXiv},
  url={https://arxiv.org/abs/2505.08751}
}

@misc{hazman2025whatmakes,
  title={What Makes a Meme a Meme? Identifying Memes for Memetics-Aware Dataset Creation},
  author={Hazman, Muzhaffar and McKeever, Susan and Griffith, Josephine},
  year={2025},
  url={https://ojs.aaai.org/index.php/ICWSM/article/view/35843}
}

@misc{hu2023tifa,
  title={TIFA: Accurate and interpretable text-to-image faithfulness evaluation with question answering},
  author={Hu, Yushi and others},
  year={2023},
  eprint={2303.11897},
  archivePrefix={arXiv},
  url={https://arxiv.org/abs/2303.11897}
}

@inproceedings{hessel-etal-2021-clipscore,
  title = {{CLIPScore}: A Reference-free Evaluation Metric for Image Captioning},
  author = {Hessel, Jack and others},
  booktitle = {Proceedings of the 2021 Conference on Empirical Methods in Natural Language Processing (EMNLP)},
  year = {2021},
  pages = {7514--7528},
  url={https://aclanthology.org/2021.emnlp-main.595}
}

@misc{romero2024cvqaculturallydiversemultilingualvisual,
  title={CVQA: Culturally-diverse Multilingual Visual Question Answering Benchmark}, 
  author={David Romero and others},
  year={2024},
  eprint={2406.05967},
  archivePrefix={arXiv},
  url={https://arxiv.org/abs/2406.05967}
}

@inproceedings{winata2025worldcuisines,
  title = {WorldCuisines: A Massive-Scale Benchmark for Multilingual and Multicultural Visual Question Answering on Global Cuisines},
  author = {Winata, Genta Indra and others},
  booktitle = {Proceedings of the 2025 NAACL},
  pages = {3242--3264},
  year = {2025},
  url={https://aclanthology.org/2025.naacl-long.167/}
}

@article{Bhalerao2025MultiAgentMM,
  title={Multi-Agent Multimodal Models for Multicultural Text to Image Generation},
  author={Parth Bhalerao and Mounika Yalamarty and Brian Trinh and Oana Ignat},
  journal={ArXiv},
  year={2025},
  volume={abs/2502.15972},
  url={https://api.semanticscholar.org/CorpusID:276575970}
}

@inproceedings{bai-etal-2025-power,
    title = "The Power of Many: Multi-Agent Multimodal Models for Cultural Image Captioning",
    author = "Bai, Longju  and
      Borah, Angana  and
      Ignat, Oana  and
      Mihalcea, Rada",
    editor = "Chiruzzo, Luis  and
      Ritter, Alan  and
      Wang, Lu",
    booktitle = "Proceedings of the 2025 Conference of the Nations of the Americas Chapter of the Association for Computational Linguistics: Human Language Technologies (Volume 1: Long Papers)",
    month = apr,
    year = "2025",
    address = "Albuquerque, New Mexico",
    publisher = "Association for Computational Linguistics",
    url = "https://aclanthology.org/2025.naacl-long.152/",
    doi = "10.18653/v1/2025.naacl-long.152",
    pages = "2970--2993",
    ISBN = "979-8-89176-189-6"
}

@inproceedings{bhatia-etal-2024-local,
    title = "From Local Concepts to Universals: Evaluating the Multicultural Understanding of Vision-Language Models",
    author = "Bhatia, Mehar  and
      Ravi, Sahithya  and
      Chinchure, Aditya  and
      Hwang, EunJeong  and
      Shwartz, Vered",
    editor = "Al-Onaizan, Yaser  and
      Bansal, Mohit  and
      Chen, Yun-Nung",
    booktitle = "Proceedings of the 2024 Conference on Empirical Methods in Natural Language Processing",
    month = nov,
    year = "2024",
    address = "Miami, Florida, USA",
    publisher = "Association for Computational Linguistics",
    url = "https://aclanthology.org/2024.emnlp-main.385/",
    doi = "10.18653/v1/2024.emnlp-main.385",
    pages = "6763--6782"
}

@inproceedings{sharma-etal-2020-semeval,
    title = "{S}em{E}val-2020 Task 8: Memotion Analysis- the Visuo-Lingual Metaphor!",
    author = {Sharma, Chhavi  and
      Bhageria, Deepesh  and
      Scott, William  and
      PYKL, Srinivas  and
      Das, Amitava  and
      Chakraborty, Tanmoy  and
      Pulabaigari, Viswanath  and
      Gamb{\"a}ck, Bj{\"o}rn},
    editor = "Herbelot, Aurelie  and
      Zhu, Xiaodan  and
      Palmer, Alexis  and
      Schneider, Nathan  and
      May, Jonathan  and
      Shutova, Ekaterina",
    booktitle = "Proceedings of the Fourteenth Workshop on Semantic Evaluation",
    month = dec,
    year = "2020",
    address = "Barcelona (online)",
    publisher = "International Committee for Computational Linguistics",
    url = "https://aclanthology.org/2020.semeval-1.99/",
    doi = "10.18653/v1/2020.semeval-1.99",
    pages = "759--773"
}

@inproceedings{tanaka-etal-2022-learning,
    title = "Learning to Evaluate Humor in Memes Based on the Incongruity Theory",
    author = "Tanaka, Kohtaro  and
      Yamane, Hiroaki  and
      Mori, Yusuke  and
      Mukuta, Yusuke  and
      Harada, Tatsuya",
    editor = "Wu, Xianchao  and
      Ruan, Peiying  and
      Li, Sheng  and
      Dong, Yi",
    booktitle = "Proceedings of the Second Workshop on When Creative AI Meets Conversational AI",
    month = oct,
    year = "2022",
    address = "Gyeongju, Republic of Korea",
    publisher = "Association for Computational Linguistics",
    url = "https://aclanthology.org/2022.cai-1.9/",
    pages = "81--93"
}

@inproceedings{10.1145/3477495.3532019,
author = {Xu, Bo and Li, Tingting and Zheng, Junzhe and Naseriparsa, Mehdi and Zhao, Zhehuan and Lin, Hongfei and Xia, Feng},
title = {MET-Meme: A Multimodal Meme Dataset Rich in Metaphors},
year = {2022},
isbn = {9781450387323},
publisher = {Association for Computing Machinery},
address = {New York, NY, USA},
url = {https://doi.org/10.1145/3477495.3532019},
doi = {10.1145/3477495.3532019},
booktitle = {Proceedings of the 45th International ACM SIGIR Conference on Research and Development in Information Retrieval},
pages = {2887–2899},
numpages = {13},
location = {Madrid, Spain},
series = {SIGIR '22}
}

@InProceedings{pmlr-v133-kiela21a,
  title = 	 {The Hateful Memes Challenge: Competition Report},
  author =       {Kiela, Douwe and Firooz, Hamed and Mohan, Aravind and Goswami, Vedanuj and Singh, Amanpreet and Fitzpatrick, Casey A. and Bull, Peter and Lipstein, Greg and Nelli, Tony and Zhu, Ron and Muennighoff, Niklas and Velioglu, Riza and Rose, Jewgeni and Lippe, Phillip and Holla, Nithin and Chandra, Shantanu and Rajamanickam, Santhosh and Antoniou, Georgios and Shutova, Ekaterina and Yannakoudakis, Helen and Sandulescu, Vlad and Ozertem, Umut and Pantel, Patrick and Specia, Lucia and Parikh, Devi},
  booktitle = 	 {Proceedings of the NeurIPS 2020 Competition and Demonstration Track},
  pages = 	 {344--360},
  year = 	 {2021},
  editor = 	 {Escalante, Hugo Jair and Hofmann, Katja},
  volume = 	 {133},
  series = 	 {Proceedings of Machine Learning Research},
  month = 	 {06--12 Dec},
  publisher =    {PMLR},
  url = 	 {https://proceedings.mlr.press/v133/kiela21a.html}
}

@inproceedings{kumar-nandakumar-2022-hate,
    title = "Hate-{CLIP}per: Multimodal Hateful Meme Classification based on Cross-modal Interaction of {CLIP} Features",
    author = "Kumar, Gokul Karthik  and
      Nandakumar, Karthik",
    editor = "Biester, Laura  and
      Demszky, Dorottya  and
      Jin, Zhijing  and
      Sachan, Mrinmaya  and
      Tetreault, Joel  and
      Wilson, Steven  and
      Xiao, Lu  and
      Zhao, Jieyu",
    booktitle = "Proceedings of the Second Workshop on NLP for Positive Impact (NLP4PI)",
    month = dec,
    year = "2022",
    address = "Abu Dhabi, United Arab Emirates (Hybrid)",
    publisher = "Association for Computational Linguistics",
    url = "https://aclanthology.org/2022.nlp4pi-1.20/",
    doi = "10.18653/v1/2022.nlp4pi-1.20",
    pages = "171--183"
}

@inproceedings{sharma-etal-2023-characterizing,
    title = "Characterizing the Entities in Harmful Memes: Who is the Hero, the Villain, the Victim?",
    author = "Sharma, Shivam  and
      Kulkarni, Atharva  and
      Suresh, Tharun  and
      Mathur, Himanshi  and
      Nakov, Preslav  and
      Akhtar, Md. Shad  and
      Chakraborty, Tanmoy",
    editor = "Vlachos, Andreas  and
      Augenstein, Isabelle",
    booktitle = "Proceedings of the 17th Conference of the European Chapter of the Association for Computational Linguistics",
    month = may,
    year = "2023",
    address = "Dubrovnik, Croatia",
    publisher = "Association for Computational Linguistics",
    url = "https://aclanthology.org/2023.eacl-main.157/",
    doi = "10.18653/v1/2023.eacl-main.157",
    pages = "2149--2163"
}

@inproceedings{DBLP:conf/emnlp/AdilazuardaMLSA24,
  author       = {Muhammad Farid Adilazuarda and
                  Sagnik Mukherjee and
                  Pradhyumna Lavania and
                  Siddhant Singh and
                  Alham Fikri Aji and
                  Jacki O'Neill and
                  Ashutosh Modi and
                  Monojit Choudhury},
  editor       = {Yaser Al{-}Onaizan and
                  Mohit Bansal and
                  Yun{-}Nung Chen},
  title        = {Towards Measuring and Modeling "Culture" in LLMs: {A} Survey},
  booktitle    = {Proceedings of the 2024 Conference on Empirical Methods in Natural
                  Language Processing, {EMNLP} 2024, Miami, FL, USA, November 12-16,
                  2024},
  pages        = {15763--15784},
  publisher    = {Association for Computational Linguistics},
  year         = {2024},
  url          = {https://doi.org/10.18653/v1/2024.emnlp-main.882},
  doi          = {10.18653/V1/2024.EMNLP-MAIN.882},
  bibsource    = {dblp computer science bibliography, https://dblp.org}
}

@inproceedings{10.1609/aaai.v39i27.35092,
author = {Mihalcea, Rada and Ignat, Oana and Bai, Longju and Borah, Angana and Chiruzzo, Luis and Jin, Zhijing and Kwizera, Claude and Nwatu, Joan and Poria, Soujanya and Solorio, Thamar},
title = {Why AI is WEIRD and shouldn't be this way: towards AI for everyone, with everyone, by everyone},
year = {2025},
isbn = {978-1-57735-897-8},
publisher = {AAAI Press},
url = {https://doi.org/10.1609/aaai.v39i27.35092},
doi = {10.1609/aaai.v39i27.35092},
booktitle = {Proceedings of the Thirty-Ninth AAAI Conference on Artificial Intelligence and Thirty-Seventh Conference on Innovative Applications of Artificial Intelligence and Fifteenth Symposium on Educational Advances in Artificial Intelligence},
articleno = {3194},
numpages = {14},
series = {AAAI'25/IAAI'25/EAAI'25}
}

@misc{liu2023visualinstruction,
  title={Visual Instruction Tuning},
  author={Liu, Haotian and Li, Chunyuan and Wu, Qingyang and Lee, Yong Jae},
  year={2023},
  eprint={2304.08485},
  archivePrefix={arXiv},
  url={https://arxiv.org/abs/2304.08485}
}

@inproceedings{deng-etal-2023-annotate,
    title = "You Are What You Annotate: Towards Better Models through Annotator Representations",
    author = "Deng, Naihao  and
      Zhang, Xinliang  and
      Liu, Siyang  and
      Wu, Winston  and
      Wang, Lu  and
      Mihalcea, Rada",
    editor = "Bouamor, Houda  and
      Pino, Juan  and
      Bali, Kalika",
    booktitle = "Findings of the Association for Computational Linguistics: EMNLP 2023",
    month = dec,
    year = "2023",
    address = "Singapore",
    publisher = "Association for Computational Linguistics",
    url = "https://aclanthology.org/2023.findings-emnlp.832/",
    doi = "10.18653/v1/2023.findings-emnlp.832",
    pages = "12475--12498"
}

@article{10.1145/3729239,
author = {Lin, Hongzhan and Luo, Ziyang and Wang, Bo and Yang, Ruichao and Ma, Jing},
title = {GOAT-Bench: Safety Insights to Large Multimodal Models through Meme-Based Social Abuse},
year = {2025},
publisher = {Association for Computing Machinery},
address = {New York, NY, USA},
issn = {2157-6904},
url = {https://doi.org/10.1145/3729239},
doi = {10.1145/3729239},
journal = {ACM Trans. Intell. Syst. Technol.},
month = apr
}

@misc{zhao2025memereacon,
  title={MemeReaCon: Probing Contextual Meme Understanding in Large Vision-Language Models},
  author={Zhao, Zhengyi and Zhang, Shubo and Zhang, Yuxi and Zhao, Yanxi and Zhang, Yifan and Wang, Zezhong and Wang, Huimin and Zhao, Yutian and Liang, Bin and Zheng, Yefeng and Li, Binyang and Wong, Kam-Fai and Wu, Xian},
  year={2025},
  eprint={2505.17433},
  archivePrefix={arXiv},
  url={https://arxiv.org/abs/2505.17433}
}

@inproceedings{muhammad-etal-2025-brighter,
    title = "{BRIGHTER}: {BRI}dging the Gap in Human-Annotated Textual Emotion Recognition Datasets for 28 Languages",
    author = "Muhammad, Shamsuddeen Hassan  and
      Ousidhoum, Nedjma  and
      Abdulmumin, Idris  and
      Wahle, Jan Philip  and
      Ruas, Terry  and
      Beloucif, Meriem  and
      de Kock, Christine  and
      Surange, Nirmal  and
      Teodorescu, Daniela  and
      Ahmad, Ibrahim Said  and
      Adelani, David Ifeoluwa  and
      Aji, Alham Fikri  and
      Ali, Felermino D. M. A.  and
      Alimova, Ilseyar  and
      Araujo, Vladimir  and
      Babakov, Nikolay  and
      Baes, Naomi  and
      Bucur, Ana-Maria  and
      Bukula, Andiswa  and
      Cao, Guanqun  and
      Tufi{\~n}o, Rodrigo  and
      Chevi, Rendi  and
      Chukwuneke, Chiamaka Ijeoma  and
      Ciobotaru, Alexandra  and
      Dementieva, Daryna  and
      Gadanya, Murja Sani  and
      Geislinger, Robert  and
      Gipp, Bela  and
      Hourrane, Oumaima  and
      Ignat, Oana  and
      Lawan, Falalu Ibrahim  and
      Mabuya, Rooweither  and
      Mahendra, Rahmad  and
      Marivate, Vukosi  and
      Panchenko, Alexander  and
      Piper, Andrew  and
      Ferreira, Charles Henrique Porto  and
      Protasov, Vitaly  and
      Rutunda, Samuel  and
      Shrivastava, Manish  and
      Udrea, Aura Cristina  and
      Wanzare, Lilian Diana Awuor  and
      Wu, Sophie  and
      Wunderlich, Florian Valentin  and
      Zhafran, Hanif Muhammad  and
      Zhang, Tianhui  and
      Zhou, Yi  and
      Mohammad, Saif M.",
    editor = "Che, Wanxiang  and
      Nabende, Joyce  and
      Shutova, Ekaterina  and
      Pilehvar, Mohammad Taher",
    booktitle = "Proceedings of the 63rd Annual Meeting of the Association for Computational Linguistics (Volume 1: Long Papers)",
    month = jul,
    year = "2025",
    address = "Vienna, Austria",
    publisher = "Association for Computational Linguistics",
    url = "https://aclanthology.org/2025.acl-long.436/",
    doi = "10.18653/v1/2025.acl-long.436",
    pages = "8895--8916"
}

@misc{kannen2024beyond,
  title={Beyond Aesthetics: Cultural Competence in Text-to-Image Models},
  author={Kannen, Nithish and Palani, Arjun and Ramaswamy, Vikram V. and Russakovsky, Olga and Fei-Fei, Li},
  year={2024},
  eprint={2407.06863},
  archivePrefix={arXiv},
  url={https://arxiv.org/abs/2407.06863}
}

@article{umuteam2024,
  title={UMUTeam at SemEval-2024 Task 4: Multilingual Detection of Persuasion Techniques in Memes},
  author={Garc{\'\i}a-D{\'\i}az, Jos{\'e} Antonio and others},
  journal={Proceedings of SemEval-2024},
  year={2024},
  url={https://aclanthology.org/2024.semeval-1.224/}
}

@misc{mgmcf2024,
  title={Multi-Granular Multimodal Clue Fusion for Meme Understanding},
  author={Zheng, Li and Wang, Xiaoming and Chen, Yuhan and Liu, Jianwei},
  year={2024},
  eprint={2503.12560},
  archivePrefix={arXiv},
  url={https://arxiv.org/abs/2503.12560}
}

@article{mutheu2023cross,
  title={Cross-Cultural Differences in Online Communication Patterns},
  author={Mutheu, Sarah},
  journal={Journal of Communication},
  volume={4},
  number={1},
  pages={31--42},
  year={2023},
  url={https://carijournals.org/journals/JCOMM/article/view/1654}
}

@misc{khanuja2024towards,
  title={Towards Automatic Evaluation for Image Transcreation},
  author={Khanuja, Simran and Iyer, Vivek and He, Claire and Neubig, Graham},
  year={2024},
  eprint={2412.13717},
  archivePrefix={arXiv},
  url={https://arxiv.org/abs/2412.13717}
}

@misc{wang2024qwen2vl,
  title={Qwen2-VL: Enhancing Vision-Language Model's Perception of the World at Any Resolution},
  author={Wang, Peng and Bai, Shuai and Tan, Sinan and Wang, Shijie and Fan, Zhihao and Bai, Jinze and Chen, Keqin and Liu, Xuejing and Wang, Jialin and Ge, Wenbin and others},
  year={2024},
  eprint={2409.12191},
  archivePrefix={arXiv},
  url={https://arxiv.org/abs/2409.12191}
}

@misc{cao2023prompthate,
  title={PromptHate: Prompting for Hateful Meme Classification},
  author={Cao, Rui and Lee, Roy Ka-Wei and Hoang, Tuan-Anh and Pang, Junmo and Kawaguchi, Kenji and Zimmermann, Roger},
  year={2023},
  eprint={2302.04156},
  archivePrefix={arXiv},
  url={https://arxiv.org/abs/2302.04156}
}

@misc{naous2023having,
  title={Having beer after prayer? Measuring cultural bias in large language models},
  author={Naous, Tarek and Ryan, Michael J and Ritter, Alan and Xu, Wei},
  year={2023},
  eprint={2305.14456},
  archivePrefix={arXiv},
  url={https://arxiv.org/abs/2305.14456}
}

\appendix

\section{Meme Transcreation Framework}\label{sec:pipeline}

Examples of visual recommendations from our Stage 1 LLaVA output to the FLUX.1 model:

\begin{enumerate}
    \item Create a cartoon image using Tom and Jerry in a detailed pose and expression. Tom, wearing his usual red shirt, is standing behind Jerry, who is dressed in his classic blue sweater. Both characters have a slightly confused look on their faces. Jerry is scratching his head while Tom looks away with a slight frown. Background: An indoor setting that resembles a cocktail party or garden tea event, with blurred figures of people in the background engaged in conversation. Style: Keep Tom and Jerry's traditional animation style with bold lines and solid colors. Mood: Soft focus and warm lighting, suggesting an evening event.

    \item Create a cartoon image using Bugs Bunny in a sitting pose, looking upward with a surprised or bewildered expression. - Background: A dimly lit room with a desk cluttered with various toys and a window showing a starry night sky. - Style: Retain the classic animation style with bold lines and vibrant colors. - Mood: Soft, nostalgic lighting with a hint of melancholy.

\end{enumerate}

\section{Dataset}

\subsection{Dataset Characteristics}\label{sec:data}

\subsection{Topic Distribution Analysis}
\label{sec:topic_analysis}
We applied enhanced weighted topic detection using Qwen-VL-Max \cite{Qwen-VL} to conduct a comprehensive topic analysis across all 6,315 filtered memes, revealing fundamental differences in cultural priorities and humor focus.

\noindent\textbf{Chinese Meme Topics.}
Table \ref{tab:chinese_topics} presents the top 10 topics in Chinese memes, collectively covering 97.2\% of the dataset.

\begin{table*}[htbp]
\centering
\small
\setlength{\tabcolsep}{4pt}
\setlength{\abovecaptionskip}{3pt}
\setlength{\belowcaptionskip}{3pt}
\caption{Topic Distribution in Chinese Memes (N=3,165)}
\label{tab:chinese_topics}
\begin{tabular}{@{}clrp{5cm}@{}}
\toprule
\textbf{\#} & \textbf{Topic} & \textbf{Count (\%)} & \textbf{Cultural Significance} \\
\midrule
1 & Internet Culture & 1,931 (61.0\%) & Digital lifestyle dominance, social media \\
2 & Technology Digital & 337 (10.6\%) & Tech adaptation, AI integration \\
3 & Work Career & 216 (6.8\%) & 996 culture, career pressure \\
4 & Social Relationships & 148 (4.7\%) & Friendships, social dynamics \\
5 & Communication Language & 126 (4.0\%) & Language barriers, expression styles \\
6 & Personality Psychology & 115 (3.6\%) & Individual traits, emotional responses \\
7 & Education Learning & 65 (2.1\%) & Academic pressure, Gaokao system \\
8 & Family Dynamics & 61 (1.9\%) & Family relationships, generational gaps \\
9 & Animals Pets & 46 (1.5\%) & Pet culture, cute content \\
10 & Entertainment Media & 32 (1.0\%) & Movies, shows, celebrity culture \\
\bottomrule
\multicolumn{4}{@{}l@{}}{\footnotesize \textit{Top 10 Total: 3,077 memes (97.2\%)}}
\end{tabular}
\vspace{-1em}
\end{table*}

\noindent\textbf{American Meme Topics.}
Table \ref{tab:american_topics} presents the top 10 topics in American memes, collectively covering 97.7\% of the dataset.

\begin{table*}[htbp]
\centering
\small
\setlength{\tabcolsep}{4pt}
\setlength{\abovecaptionskip}{3pt}
\setlength{\belowcaptionskip}{3pt}
\caption{Topic Distribution in American Memes (N=3,150)}
\label{tab:american_topics}
\begin{tabular}{@{}clrp{5cm}@{}}
\toprule
\textbf{\#} & \textbf{Topic} & \textbf{Count (\%)} & \textbf{Cultural Significance} \\
\midrule
1 & Internet Culture & 1,651 (52.4\%) & Social media, viral content, online trends \\
2 & Technology Digital & 475 (15.1\%) & Tech innovation, digital lifestyle \\
3 & Education Learning & 247 (7.8\%) & School experiences, college culture \\
4 & Work Career & 198 (6.3\%) & Job market, work-life balance \\
5 & Family Dynamics & 155 (4.9\%) & Family relationships, parenting \\
6 & Communication Language & 87 (2.8\%) & Expression styles, conversation humor \\
7 & Gaming Entertainment & 85 (2.7\%) & Video games, gaming culture, esports \\
8 & Personality Psychology & 67 (2.1\%) & Individual psychology, personality types \\
9 & Social Relationships & 57 (1.8\%) & Friendships, social interactions \\
10 & Entertainment Media & 54 (1.7\%) & Movies, TV shows, celebrity content \\
\bottomrule
\multicolumn{4}{@{}l@{}}{\footnotesize \textit{Top 10 Total: 3,076 memes (97.7\%)}}
\end{tabular}
\vspace{-1em}
\end{table*}

\noindent\textbf{Cross-Cultural Topic Comparisons.}
Table \ref{tab:topic_comparison} highlights key differences in topic priorities between the two cultures.

\begin{table*}[htbp]
\centering
\small
\setlength{\abovecaptionskip}{3pt}
\setlength{\belowcaptionskip}{3pt}
\caption{Cross-Cultural Topic Priority Comparison}
\label{tab:topic_comparison}
\begin{tabular}{@{}lccp{5.5cm}@{}}
\toprule
\textbf{Topic} & \textbf{CN} & \textbf{US} & \textbf{Interpretation} \\
\midrule
Internet Culture & \#1 (61.0\%) & \#1 (52.4\%) & Both dominant; China more concentrated \\
Technology Digital & \#2 (10.6\%) & \#2 (15.1\%) & US higher tech innovation focus \\
Education Learning & \#7 (2.1\%) & \#3 (7.8\%) & US: daily life; China: high-stakes pressure \\
Family Dynamics & \#8 (1.9\%) & \#5 (4.9\%) & US: frequent topic; China: serious element \\
Gaming Entertainment & -- & \#7 (2.7\%) & US leisure vs. China work/study priority \\
Work Career & \#3 (6.8\%) & \#4 (6.3\%) & Similar priority, different intensity \\
\bottomrule
\end{tabular}
\vspace{-1em}
\end{table*}

Key cultural patterns revealed: (1) \textit{Digital Concentration}—Chinese memes more heavily focused on internet/digital life (71.6\% combined vs. 67.5\% in US); (2) \textit{Educational Values}—American memes treat education as casual daily experience (7.8\%), Chinese memes reflect intense academic pressure (2.1\%); (3) \textit{Family Representation}—American memes more frequently feature family humor (4.9\%) vs. Chinese hierarchical respect (1.9\%); (4) \textit{Leisure vs. Achievement}—American gaming culture prominent (2.7\%), absent from Chinese top 10.

\subsection{Emotion Distribution Analysis}
\label{sec:emotion_analysis}
Using Qwen-based \cite{Qwen-VL} automated emotion analysis, we classified all 6,315 memes according to Ekman's six basic emotions, providing insights into cross-cultural emotional expression patterns.

\noindent\textbf{Chinese Meme Emotions.}
Table \ref{tab:chinese_emotions} presents the emotion distribution in Chinese memes.

\begin{table*}[htbp]
\centering
\small
\setlength{\abovecaptionskip}{3pt}
\setlength{\belowcaptionskip}{3pt}
\caption{Emotion Distribution in Chinese Memes (N=3,165)}
\label{tab:chinese_emotions}
\begin{tabular}{@{}lrp{7cm}@{}}
\toprule
\textbf{Emotion} & \textbf{Count (\%)} & \textbf{Cultural Context} \\
\midrule
Joy & 2,193 (69.3\%) & Dominant positive humor expression \\
Anger & 263 (8.3\%) & Frustration, social critique \\
Sadness & 258 (8.2\%) & Melancholy, disappointment \\
Surprise & 213 (6.7\%) & Shock, unexpected situations \\
Fear & 144 (4.5\%) & Anxiety, worry \\
Disgust & 94 (3.0\%) & Revulsion, distaste \\
\bottomrule
\end{tabular}
\vspace{-1em}
\end{table*}

\noindent\textbf{American Meme Emotions.}
Table \ref{tab:american_emotions} presents the emotion distribution in American memes.

\begin{table*}[htbp]
\centering
\small
\setlength{\abovecaptionskip}{3pt}
\setlength{\belowcaptionskip}{3pt}
\caption{Emotion Distribution in American Memes (N=3,150)}
\label{tab:american_emotions}
\begin{tabular}{@{}lrp{7cm}@{}}
\toprule
\textbf{Emotion} & \textbf{Count (\%)} & \textbf{Cultural Context} \\
\midrule
Joy & 2,325 (73.8\%) & Primary emotional expression \\
Fear & 219 (7.0\%) & Anxiety, relatable worries \\
Anger & 217 (6.9\%) & Frustration, social commentary \\
Surprise & 148 (4.7\%) & Unexpected, absurd humor \\
Disgust & 140 (4.4\%) & Cringe, distasteful situations \\
Sadness & 101 (3.2\%) & Disappointment, darker humor \\
\bottomrule
\end{tabular}
\vspace{-1em}
\end{table*}

\noindent\textbf{Cross-Cultural Emotion Comparisons.}
Table \ref{tab:emotion_comparison} highlights key differences in emotional expression priorities.

\begin{table*}[htbp]
\centering
\small
\setlength{\abovecaptionskip}{3pt}
\setlength{\belowcaptionskip}{3pt}
\caption{Cross-Cultural Emotion Priority Comparison}
\label{tab:emotion_comparison}
\begin{tabular}{@{}lccp{5.5cm}@{}}
\toprule
\textbf{Emotion} & \textbf{CN} & \textbf{US} & \textbf{Interpretation} \\
\midrule
Joy & \#1 (69.3\%) & \#1 (73.8\%) & Both dominant; US slightly higher positivity \\
Anger & \#2 (8.3\%) & \#3 (6.9\%) & China: more direct frustration expression \\
Sadness & \#3 (8.2\%) & \#6 (3.2\%) & China: 2.5× higher melancholic acceptance \\
Surprise & \#4 (6.7\%) & \#4 (4.7\%) & Similar priority, China higher absurdist humor \\
Fear & \#5 (4.5\%) & \#2 (7.0\%) & US: anxiety culture, relatable worry themes \\
Disgust & \#6 (3.0\%) & \#5 (4.4\%) & US: higher cringe/distaste expression \\
\bottomrule
\end{tabular}
\vspace{-1em}
\end{table*}

Key emotional patterns: (1) \textit{Positive Emphasis}—Both cultures prioritize joy, Americans showing slightly higher positive focus (73.8\% vs. 69.3\%); (2) \textit{Sadness Acceptance}—Chinese memes express sadness 2.5× more frequently, reflecting cultural acceptance of melancholic humor; (3) \textit{Anxiety Expression}—American memes emphasize fear-based content (7.0\% vs. 4.5\%), aligning with therapeutic humor trends; (4) \textit{Anger Manifestation}—Chinese memes show higher anger (8.3\% vs. 6.9\%), possibly reflecting more direct emotional expression; (5) \textit{Cringe Culture}—American memes display higher disgust representation (4.4\% vs. 3.0\%), consistent with cringe comedy trends.

\noindent\textbf{Impact of Joy Dominance on Experimental Design.}
The overwhelming dominance of Joy in both datasets (69.3\% Chinese, 73.8\% American) has important implications for our transcreation experiments and evaluation: (1) \emph{Positive Bias in Evaluation}—Since most transcreated memes will naturally preserve joyful emotions, our system may appear more successful at humor preservation simply due to the high baseline of positive content. This necessitates careful interpretation of intent preservation scores in Chapter~\ref{ch:evaluation}; (2) \emph{Limited Negative Emotion Testing}—With less than 30\% of memes expressing negative emotions (anger, sadness, fear, disgust), our system receives limited training signals for adapting complex negative emotional tones across cultures, potentially underrepresenting challenges in transcreating emotionally nuanced content; (3) \emph{Generalizability Concerns}—The skewed distribution means our findings may generalize better to lighthearted, positive meme content than to darker, satirical, or critical humor styles; (4) \emph{Cultural Authenticity vs. Emotional Consistency}—The Joy dominance simplifies one aspect of transcreation (emotional tone transfer) while placing greater emphasis on cultural reference adaptation as the primary challenge. Despite this limitation, the Joy-dominant distribution accurately reflects real-world meme ecosystems where positive, shareable content naturally dominates social media platforms—making our experimental conditions ecologically valid even if not emotionally balanced.

\section{Evaluation Metrics}\label{sec:metrics}
\noindent\textbf{Quantitative Dimensions (1-5 scale):}
\begin{description}[nosep]
    \item[Caption Quality] Evaluates whether the generated caption works effectively as meme text, considering clarity, readability, appropriate meme language/tone, engaging phrasing, and proper text formatting.
    
    \item[Image Quality] Assesses whether the generated image functions effectively as a meme visual, considering visual clarity and quality, appropriate meme composition, recognizable elements/characters, and visual appeal and memorability.
    
    \item[Synergy] Measures how well image and caption work together, evaluating coherent message delivery, emotional or humorous impact, logical connection between visual and text, and overall meme effectiveness.
    
    \item[Cultural Fit] Evaluates cultural adaptation quality, including alignment with target culture's humor style, appropriate cultural references, target audience relatability, and avoidance of cultural misunderstandings.
    
    \item[Intent Preservation] Assesses preservation of the original meme's intent, including message consistency, emotional tone preservation, humorous effect maintenance, and core meaning retention.
    
    \item[Overall Score] The average of all dimension scores and reflects overall quality.
\end{description}

\section{Human Evaluator Profiles}
\label{sec:evaluators}

Our evaluation employed three bilingual, bicultural evaluators with a deep understanding of both Chinese and US cultural contexts:

\textbf{Evaluator 1.} Native Chinese speaker with 10+ years US residence, PhD in Communication Studies. Regular engagement with both Weibo/Xiaohongshu and Reddit meme communities. Assessment style: Entertainment-focused, generous scoring emphasizing humor effectiveness over technical perfection. Mean overall score: 4.42/5.0.

\textbf{Evaluator 2.} American-born Chinese with native-level proficiency in Mandarin and US, MA in Comparative Cultural Studies. Active participation in both Chinese and US digital cultures. Assessment style: Balanced and objective, applying consistent standards across dimensions. Mean overall score: 4.09/5.0. Showed highest correlation with Qwen-VL-Max (F1 = 0.925, $r = 0.964$).

\textbf{Evaluator 3.} Native Chinese speaker with 12+ years of US experience, professional translator with meme localization background. Expertise in cultural adaptation nuances. Assessment style: Critical and quality-focused, emphasizing cultural authenticity and linguistic precision. Mean overall score: 3.31/5.0.

All evaluators received identical structured prompts specifying six evaluation dimensions, worked independently without access to others' ratings, and maintained consistency through detailed scoring rubrics. Inter-evaluator correlations demonstrate moderate to strong agreement: Evaluator 1-2 ($r = 0.72$), Evaluator 1-3 ($r = 0.58$), Evaluator 2-3 ($r = 0.81$), confirming reliable yet stylistically distinct evaluation perspectives.

\section{VLM Evaluator Details}
\label{sec:llm_details}

Six VLMs served as automated evaluators, selected for multilingual (Chinese-English) capability, multi-image processing, and reproducibility:

\begin{itemize}
\item \textbf{Qwen-VL-Max} (Alibaba Cloud): Commercial API with extensive Chinese-English training, demonstrated exceptional human correlation ($r = 0.926$).
\item \textbf{LLaVA-v1.6-Vicuna-13B}: Same architecture as transcreation system, showed no meaningful correlation ($r = 0.005$).
\item \textbf{InternVL3-8B/14B}: Recent open-source models with strong vision capabilities, achieved weak positive correlation (8B: $r = -0.049$, 14B: $r = 0.263$).
\item \textbf{Qwen3-VL-8B-Instruct}: Smaller Qwen variant, weak correlation ($r = 0.252$).
\item \textbf{Aya-vision-8b}: Massively multilingual model, slight negative correlation ($r = -0.043$).
\end{itemize}

All LLMs received identical prompts specifying evaluation dimensions and rating scales. Temperature set to 0.7 for balanced consistency-creativity tradeoff.

\section{Transcreation Prompts}
\label{sec:prompts}

\textbf{Stage 1 (LLaVA 1.6) - Cultural Analysis Prompt Example:}

\begin{quote}
\small
\textit{You are a cultural adaptation expert. Analyze this [SOURCE CULTURE] meme and create a transcreated version for [TARGET CULTURE] audiences. Your response should include:}

\textit{1. Cultural Context Analysis: Identify culture-specific elements (references, idioms, visual symbols, humor mechanisms)}

\textit{2. Intent Extraction: What is the core message/emotion/joke?}

\textit{3. Target Culture Mapping: Find equivalent concepts in [TARGET CULTURE]}

\textit{4. Transcreated Caption: Generate a new caption preserving intent while using [TARGET CULTURE] appropriate references and style}

\textit{5. Visual Recommendations: Describe ideal visual template (characters, setting, composition) culturally appropriate for [TARGET CULTURE]}
\end{quote}

\textbf{Stage 2 (FLUX.1) - Visual Generation Prompt Example:}

\begin{quote}
\small
\textit{Create a meme-style image: [LLaVA's visual recommendations]. Style: internet meme, high contrast, recognizable characters, clear composition suitable for text overlay. [TARGET CULTURE]-appropriate visual elements. Resolution: 1024x1024px.}
\end{quote}

Full prompt templates with examples will be made available in our public repository upon acceptance.

\section{Example Transcreations}
\label{sec:examples}

\textbf{Success Example - US$\rightarrow$Chinese:}

\textit{Source (US):} "Nobody: Absolutely nobody: Me at 3 am:" [Image: Person raiding refrigerator]

\textit{Transcreated (Chinese):} \zh{"深夜两点的我:"} (Me at 2 am) [Image: Cartoon cat staring at food] 

\textit{Adaptation rationale:} Replaced human figure with animal imagery (preferred in Chinese memes), adjusted time (2 am vs 3 am reflects Chinese sleep patterns), simplified narrative structure for conciseness.

\textbf{Challenge Example - Chinese$\rightarrow$US:}

\textit{Source (Chinese):} \zh{"内卷"} (involution) concept with study-exhausted imagery

\textit{Transcreated (US):} "The grind never stops" [Office worker imagery]

\textit{Limitation:} US lacks a precise equivalent for \zh{"内卷"} (intensifying competition in zero-sum environments). "Grind culture" captures work intensity but misses the systemic competition aspect, illustrating cultural untranslatability challenges.

Additional examples and failure case analysis are available in the supplementary materials.

\begin{figure}[h]
\centering
\includegraphics[width=0.9\columnwidth]{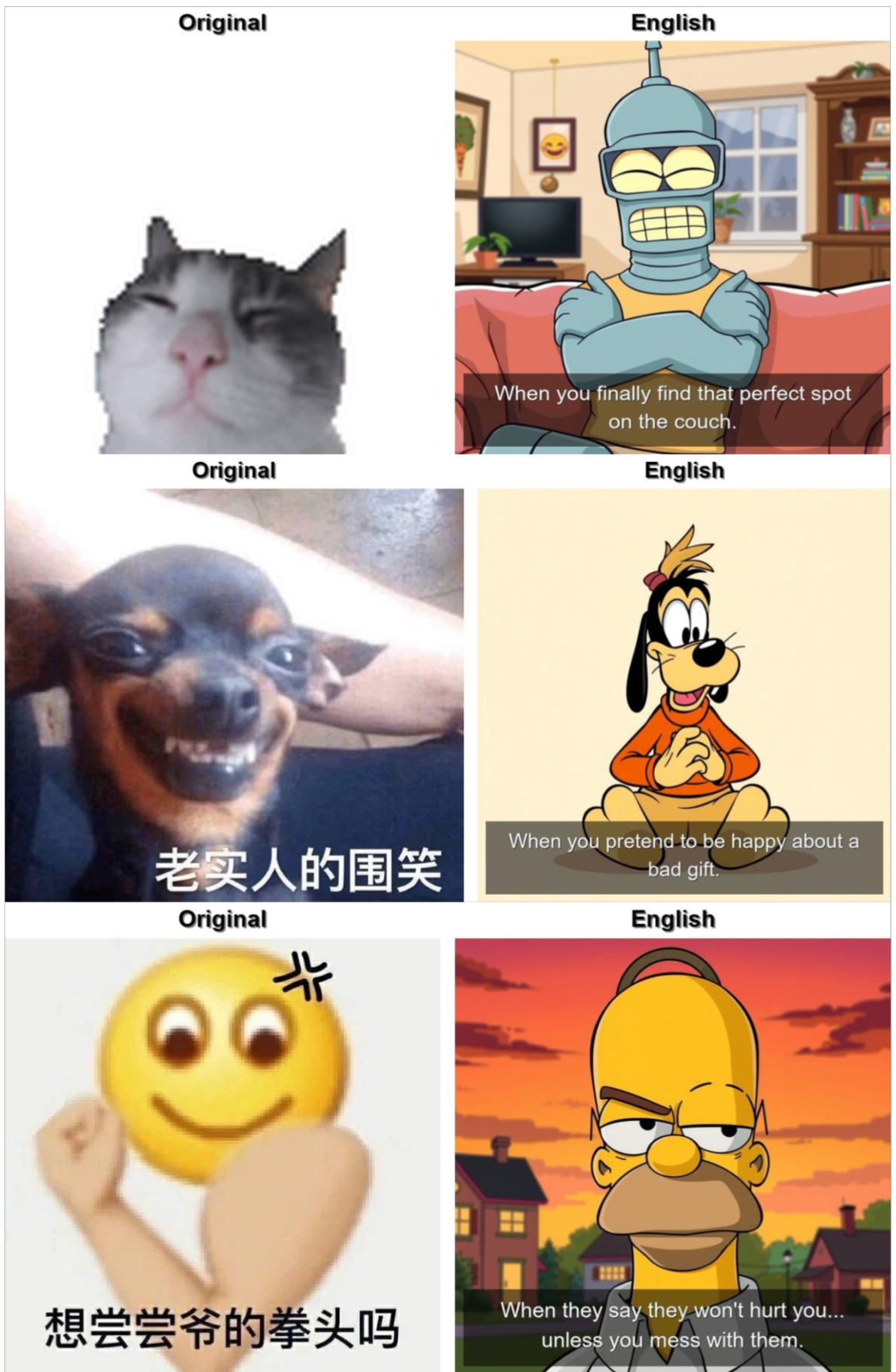}
\caption{Examples of successful Chinese$\rightarrow$US meme transcreations. \\\textbf{\textcolor{emphblue}{Original (middle left — dog meme):}} 
    ``Honest smile''\\\textbf{\textcolor{emphblue}{Original (bottom left — angry emoji meme):}} 
    ``You looking for a knuckle sandwich?''}
\label{fig:chinese_to_us_examples_app}
\end{figure}

\begin{figure}[h]
\centering
\includegraphics[width=0.9\columnwidth]{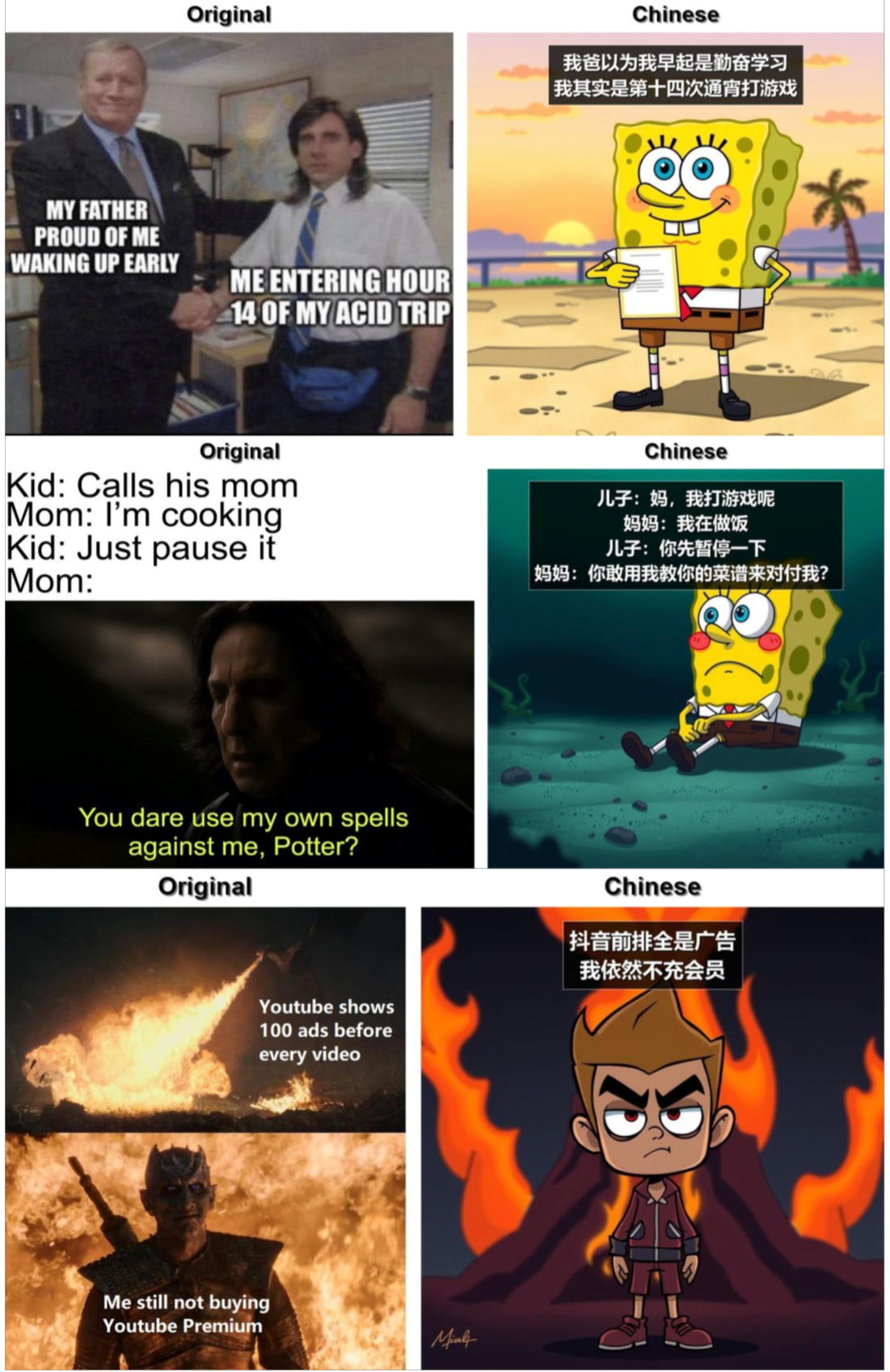}
\caption{Examples of successful US$\rightarrow$Chinese meme transcreations. \\\textbf{\textcolor{emphblue}{Transcreated (top right — spongebob meme):}} 
    ``Dad thinks I'm up early studying, I'm really on my 14th straight gaming session''\\\textbf{\textcolor{emphblue}{Transcreated (middle right — spongebob meme):}} 
    ``Kid: Mom, I'm playing a game  \\Mom: I'm cooking  \\Kid: Can you pause it? \\Mom: How dare you use my own teachings against me?''\\\textbf{\textcolor{emphblue}{Transcreated (middle right — Spencer Wright meme):}} 
    ``Tiktok be spamming ads upfront, still not gonna pay for premium''}
\label{fig:us_to_chinese_examples_app}
\end{figure}
\end{document}